\documentclass[pra,twocolumn,superscriptaddress,showpacs,amsmath,amstex,amssymb,citeautoscript]{revtex4-1}

\usepackage[T1]{fontenc}
\usepackage[utf8]{inputenc}
\usepackage{lipsum}
\usepackage{amsmath}
\usepackage{amssymb}
\usepackage{bbm}
\usepackage{braket}
\usepackage{xcolor}
\usepackage{pifont}
\usepackage[mathscr]{euscript}
\usepackage[shortlabels]{enumitem}
\usepackage{graphicx}
\usepackage{lipsum}
\allowdisplaybreaks
\usepackage{float}
\usepackage{graphicx}
\usepackage{dsfont}
\usepackage{comment}
\usepackage[colorlinks=true]{hyperref}  
\hypersetup{
    bookmarks=true,         
    unicode=false,          
    pdftoolbar=true,        
    pdfmenubar=true,        
    pdffitwindow=false,     
    pdfstartview={FitH},    
    pdftitle={HK models for unconventional magnets},    
    pdfauthor={Rickelt, Sedov, Scheurer},     
    pdfsubject={},   
    pdfcreator={},   
    pdfproducer={}, 
    pdfkeywords={} {} {}, 
    pdfnewwindow=true,      
    colorlinks=true,       
    linkcolor=blue, 
    citecolor=blue,        
    filecolor=magenta,      
    urlcolor=blue           
}

\newcommand{\equref}[1]{Eq.~(\ref{#1})}

\newcommand{\figref}[1]{Fig.~\ref{#1}}
\newcommand{\refcite}[1]{Ref.~\onlinecite{#1}}

\newcommand{\pd}{{\phantom{\dagger}}}
\newcommand{\diff}{\mathrm{d}}

\renewcommand{\vec}[1]{\boldsymbol{#1}}

\definecolor{wrongultramarine}{rgb}{1,0.5,0}

\begin{document}

\title{Hatsugai-Kohmoto-like Models for Altermagnets and Odd-Parity Magnets}

\author{Konstantin Rickelt}
\thanks{These two authors contributed equally}
\affiliation{Institute for Theoretical Physics III, University of Stuttgart, 70550 Stuttgart, Germany}

\author{Denis Sedov}
\thanks{These two authors contributed equally}
\affiliation{Institute for Theoretical Physics III, University of Stuttgart, 70550 Stuttgart, Germany}

\author{Mathias S.~Scheurer}
\affiliation{Institute for Theoretical Physics III, University of Stuttgart, 70550 Stuttgart, Germany}

\begin{abstract}
We introduce a generalized Hatsugai-Kohmoto multi-orbital model and study its phase diagram and physical properties in the additional presence of perturbations that lift any extensive ground-state degeneracies. The unperturbed, exactly solvable model already displays a rich set of spectral functions, including regimes reminiscent of unconventional magnets. We map the first-order study of additional spatially local multi-orbital Hubbard interactions to a Heisenberg model in momentum space, which leads to symmetry-breaking instabilities already at weak coupling. Interestingly, translational-symmetry breaking orders, such as antiferromagnetism, are excluded. Instead, in addition to ferromagnetism, unconventional $p$-wave and $d$-wave magnets occur, characterized by spin order on the bonds of the underlying square lattice. Adding another type of momentum-space interaction, which still allows to solve the model exactly, is shown to stabilize a non-degenerate singlet ground state that retains the spin splitting characteristic of unconventional magnets. We discuss its impact on the spin structure factor. Taken together, our findings show that Hatsugai-Kohmoto-like models provide a rich playground for unconventional magnetism.
\end{abstract}

\maketitle
The impact of strong interactions and symmetry breaking in itinerant electronic systems is one of the most challenging problems of quantum many-body physics but at the same time associated with remarkably rich phenomenology \cite{RevModPhys.82.2421,RevModPhys.75.913,ReviewPnictides,Santiago_2017}. In that regard, the Hastugai-Kohmoto (HK) model \cite{HKModelOriginal,Zhao_2025}, which is local in momentum space, provides an interesting route to analyzing Mott and non-Fermi liquid physics in a fine-tuned and spatially non-local, yet solvable way \cite{CONTINENTINO1994619,PhysRevB.57.1340,PhysRevB.61.7941}. The interest in the HK and closely related models, with modifications in the quadratic part \cite{PhysRevB.108.165136,PhysRevB.111.045126,2025arXiv251104141B,PhysRevResearch.6.033235}, such as spin-orbit coupling \cite{PhysRevB.108.035121,PhysRevB.109.075147} or magnetism/magnetic fields \cite{PhysRevB.108.035121,2025arXiv251104141B} and periodic Anderson models \cite{PhysRevB.106.155119,PhysRevB.108.195145,PhysRevB.109.064517}, or where each momentum is coupled to finitely many other momenta \cite{PhysRevLett.133.166501,TwistedHubbard,2026arXiv260406588M}, has recently been revived significantly; for instance, HK-like models have been employed to study the interplay of correlations with pairing \cite{PPHKModel,PhysRevB.105.184509,Li_2022,PhysRevB.109.064517,PhysRevB.111.075115,Zhu_2021,CORSINO2025131070,2025arXiv251017452C} or topology \cite{PPTopology,PhysRevResearch.5.013162,PhysRevB.108.035121,PhysRevLett.131.106601,PhysRevB.108.195145,PhysRevResearch.6.033235,PhysRevB.109.075147,PhysRevB.109.165129}, to address Fermi arcs \cite{PhysRevLett.133.166501,PhysRevB.103.024529}, and beyond \cite{PhysRevD.99.094030,2024MPLB...3850027Z,1qzc-6c6n,PhysRevB.111.075124,r4q9-hm45,PhysRevB.109.165129,Zhao_2023,PerturbationTheory,PhysRevResearch.6.L032018}. 

In this work, we start from a two-orbital HK-like model, where---similar to what was dubbed ``orbital HK models’’ in \refcite{PhysRevB.108.165136}---the interaction and the kinetic terms do not commute. However, we still obtain parameter regimes with extensive ground-state degeneracies in momentum-space regions that are shaped by the interplay of interactions and hopping processes. The associated correlation-driven changes in the spectral functions can be similar to those of unconventional magnets, another recent development in the field of itinerant magnetism \cite{PhysRevX.12.040501,magnetism5030017,2025arXiv251009170L,ReviewUnconventionalMagnetism}. Treating local interactions as a perturbation, we also find that the two major classes of unconventional magnets, altermagnets and odd-parity magnets, naturally arise. Besides long-range order, we consider an additional interaction term that does not spoil the exact solvability and stabilizes a unique ground state. This shows how momentum-dependent splittings of the spectral function, reminiscent of unconventional magnets, and in the absence of long-range order or extensive ground-state degeneracies can be obtained in a theoretically controlled way.

\begin{figure}[t] 
    \centering
    \includegraphics[width=0.9\columnwidth]{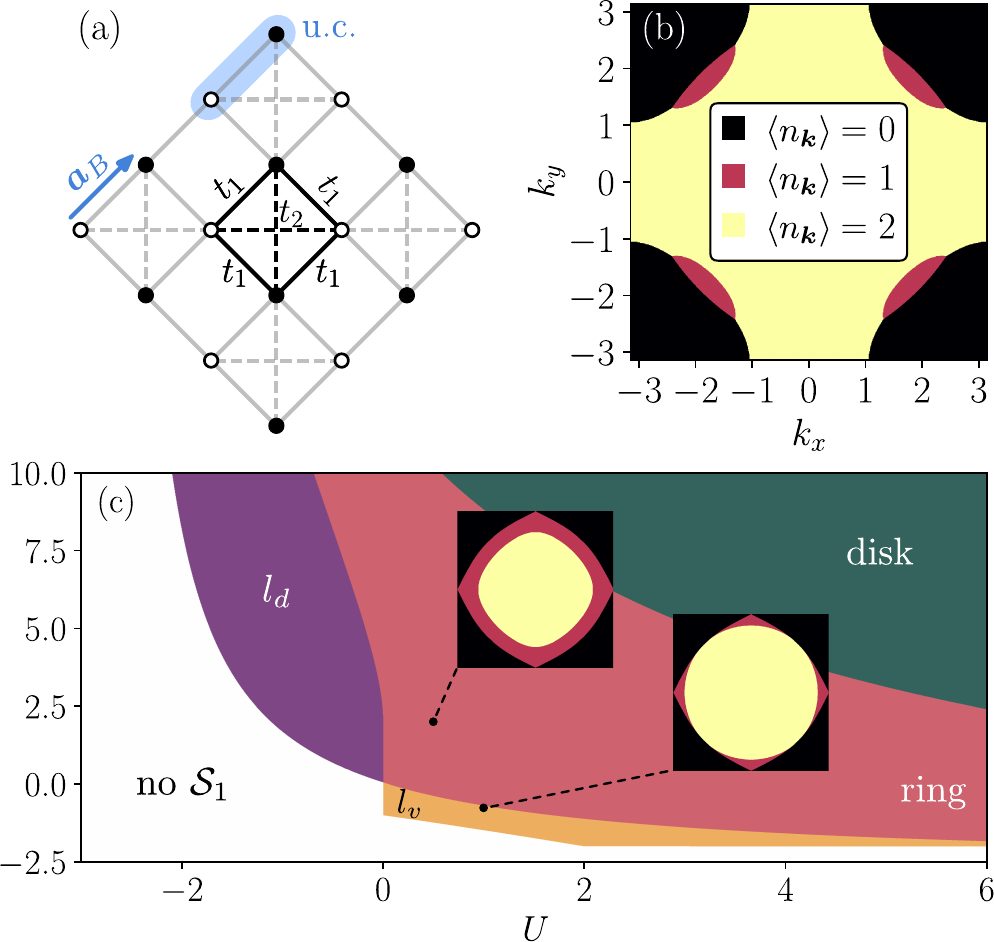}
    \caption{\textbf{(a)} Checkerboard lattice with two sublattices ($\circ=A$; $\bullet=B$), nearest ($t_1$) and next-nearest neighbor hopping ($t_2$). \textbf{(b)} Momentum-space occupation of the ground state of $H_{\mathrm{HK}}$ in the $l_d$ regime, for $U=-1.0$, $U^\prime = 4.0$. \textbf{(c)} Phase diagram of the different ground state occupation configurations  $l_d$, $l_v$ and ring (see main text and insets for definition). We took $\mu=-1.0$ and $t_2=0.5$.}
    \label{fig:HKSetup}
\end{figure}

\vspace{1em}
\textit{Model and ground state}---We consider a two-orbital HK model on the checkerboard lattice with two sublattices $l = A,B$ shown in~\figref{fig:HKSetup}a. Denoting the annihilation operator of an electron on the sublattice $l$ with spin $\sigma$ and momentum $\vec{k}$ by $c_{\vec{k},l,\sigma}$, the Hamiltonian reads
\begin{align}
    \begin{aligned}
        H_{\mathrm{HK}} &=\sum_{\boldsymbol{k},l,l',\sigma} c^\dagger_{\boldsymbol{k},l,\sigma} h_{\vec{k}}^{l,l'} c^{\pd}_{\boldsymbol{k},l^\prime,\sigma} \\
        &+ U \sum_{\boldsymbol{k},l} n_{\boldsymbol{k},l,\uparrow} n_{\boldsymbol{k},l,\downarrow}
        + U^\prime \sum_{\boldsymbol{k},\sigma,\sigma'} n_{\boldsymbol{k},A,\sigma} n_{\boldsymbol{k},B,\sigma^\prime}~, \label{eq:HK_checkerboard}
    \end{aligned}
\end{align}
where $h_{\vec{k}} = p_{\vec{k}} (\tau_0 + \tau_z) + g_{\vec{k}}(\tau_0 - \tau_z) + [r_{\vec{k}}(\tau_x - i\tau_y) + \text{H.c.}] - \mu \tau_0$ is the Bloch Hamiltonian, $p_{\vec{k}} = -t_{2} \cos(k_x), g_{\vec{k}}=-t_{2}\cos(k_y), r_{\vec{k}} = -2t_{1}\cos(k_x/2)\cos(k_y/2)$; $\tau_j$ are Pauli matrices in the sublattice space. We will set $t_1 = 1$ in the following. HK models are characterized by the full decoupling of different momenta, a feature reminiscent of non-interaction systems. In each $\vec{k}$-sector, the Hilbert space of the considered model is $2^4$-dimensional, which allows for exact diagonalization. Furthermore, due to the conservation of the number of particles, $[H_{\mathrm{HK}}, n_{\vec{k}}] = 0$, $n_{\vec{k}} = \sum_{l,\sigma}n_{\vec{k},l,\sigma}$, the closed-form expressions for eigenenergies can be found analytically \cite{SIref}.

We choose $t_2,\mu$ such that the occupation of the ground state satisfies $ \braket{n_{\vec{k}}} \in \{0,1,2\}$, for all $\vec{k}$, while occupations of $3$ and $4$ do not occur; we denote the regions of the Brillouin zone (BZ) with $m$ electrons by $\mathcal{S}_m$. In $\mathcal{S}_{0,2}$, the ground state is non-degenerate, while the singly-occupied region $\mathcal{S}_1$ is spin-degenerate which leads to an exponential ($2^{|\mathcal{S}_1|}$-fold) degeneracy of the ground state. Each of the ground states can be parametrized by the spin-configuration as follows
\begin{align}
    \ket{\{ \sigma_{\vec{k}} \}} = \prod_{\vec{k}\in\mathcal{S}_1} f^\dagger_{\vec{k},\sigma_{\vec{k}}} \ket{\mathcal{S}_2}, \label{eq:HK_degenerate_gs}
\end{align}
where $f^\dagger_{\vec{k},\sigma}$ is the creation operator of the lower one-particle band of the Bloch Hamiltonian $h_{\vec{k}}$, and $\ket{\mathcal{S}_2}$ corresponds to the non-degenerate doubly occupied region in the BZ. 

Depending on $U$ and $U'$, $\mathcal{S}_1$ assumes different shapes. It can be a simply connected filled disk (when $\mathcal{S}_2$ is a null set), ring or consist of four separated lobes. The lobes are located either along the $\Gamma \mathrm{M}$ path as shown in~\figref{fig:HKSetup}b, to which we will refer as $l_d$, or along $\Gamma \mathrm{X}$ (see inset of~\figref{fig:HKSetup}c)---denoted by $l_v$. The occupation of the ground state for the ring- and disk-shaped $\mathcal{S}_1$ resembles the occupation of the ferromagnet, whilst $l_{v/d}$-regimes share similarities with unconventional magnets. Figure~\ref{fig:HKSetup}c presents the phase diagram of the ground-state occupation. Having established the properties of the degenerate HK model ground states, we will next study different perturbations lifting the degeneracy and discuss the resulting physical properties.

\vspace{1em}
\textit{Long-range order}---We will first explore how a conventional, spatially local perturbation can lead to spin-rotational symmetry breaking and magnetic long-range order. Specifically, we consider a Hubbard-like Hamiltonian with on-site $V$ and nearest-neighbors $V'$ interaction which reads
\begin{align} 
    \Delta H &= V \sum_{\vec{R},l} n_{\vec{R}+\vec{a}_{l},\uparrow}  n_{\vec{R}+\vec{a}_{l},\downarrow} + \frac{V'}{4} \sum_{\vec{R},\vec{\delta}_j,\sigma,\sigma'} n_{\vec{R},\sigma} n_{\vec{R}+\vec{\delta}_j,\sigma'} \nonumber \\
    =&\frac{1}{N} \sum_{\vec{q},l,l',\sigma,\sigma'} \mathcal{V}_{\vec{q}}^{l,l'} : \rho_{\vec{q},l,\sigma} \rho_{-\vec{q},l',\sigma'} : ,  \label{eq:deltaH}
\end{align}
where $\vec{a}_{A}=(0,0), \vec{a}_B = (1/2,1/2)$ are $A$, $B$ sites positions in the unit cell, $\vec{\delta}_j = C_{4z}^j \vec{a}_B$ are four vectors connecting nearest $A$ and $B$ sites, cf.~\figref{fig:HKSetup}a; $\rho_{\vec{q},l,\sigma} = \sum_{\vec{k}} c^{\dagger}_{\vec{k}+\vec{q},l,\sigma} c^{\pd}_{\vec{k},l,\sigma}$; $\mathcal{V}_{\vec{q}}^{AA} = \mathcal{V}_{\vec{q}}^{BB} = V/2$, $\mathcal{V}_{\vec{q}}^{AB} = \mathcal{V}_{\vec{q}}^{BA} = V'/2 \cos(q_x/2) \cos(q_y/2)$. We will see how the bare HK ground states act as a precursor for unconventional magnetism induced by the perturbation~\equref{eq:deltaH}. One can also think of the following analysis as a perturbative solution of the deformed Hubbard model, in which the $\vec{k}$-local part is enhanced artificially to dominate the non-local contributions.
Motivated by this picture, we choose $V = \lambda U$ and $V' = \lambda U'$ with $\lambda$ being the control parameter.

Treating $\Delta H$ via first order perturbation theory, the ground state of $H_{\mathrm{HK}} + \Delta H$ is obtained by diagonalizing $P \Delta H P$, where $P= \sum_{\ket{\{ \sigma_{\vec{k}} \}}} \ket{\{ \sigma_{\vec{k}} \}}\bra{\{ \sigma_{\vec{k}} \}}$ is the projector onto the degenerate HK ground state. This subspace, characterized by fixed occupation at each momentum, can be thought of as the flat band for $\vec{k}$-local spin-flip excitations within $\mathcal{S}_1$. Therefore, we naturally find that the projected Hamiltonian contains only the effective exchange part of~\equref{eq:deltaH} which can be mapped onto the following quantum spin-1/2 Heisenberg model in momentum space \cite{SIref}
\begin{align}\begin{split}
    P \Delta H P \mapsto & \sum_{\vec{k},\vec{k}' \in \mathcal{S}_1,\vec{k}\neq\vec{k}'} -J_{\vec{k},\vec{k}'}\boldsymbol{\sigma}_{\vec{k}} \cdot \boldsymbol{\sigma}_{\vec{k}'},\\
    &J_{\vec{k},\vec{k}'} = \frac{1}{2N} \sum_{l,l'} \mathcal{V}^{l,l'}_{\vec{k} - \vec{k}'} \phi_{\vec{k},l}^* \phi_{\vec{k},l'}^{\pd} \phi_{\vec{k}',l'}^* \phi_{\vec{k}',l}^{\pd}, \label{EmergentSpinModel}
\end{split}\end{align}
where $\vec{\sigma}_{\vec{k}} = (\sigma_{\vec{k}}^x, \sigma_{\vec{k}}^y, \sigma_{\vec{k}}^z)$ are Pauli matrices, and $\phi_{\vec{k},l}$ are the wavefunctions of the lower band of $h_{\vec{k}}$ (see also \cite{PhysRevB.103.024529} where a different HK model was studied). This mapping holds for an arbitrary HK model and interaction of the form of~\equref{eq:deltaH} as long as the degeneracy of the ground state is determined only by the spin-degenerate region $\mathcal{S}_1$ of the BZ. In general, the obtained Heisenberg model is frustrated and dense, which could induce highly-entangled ground states, including states without magnetic order. This, however, will be our focus for the second type of perturbation studied below, and we will therefore use mean-field (MF) theory here to demonstrate the emergence of magnetic phases.
Limiting the discussion to collinear states, the problem of finding the MF ground state of the Hamiltonian reduces to the minimization of the classical Ising model, $E[\{\sigma_{\vec{k}}  = \pm  \}] = -  \sum_{\vec{k},\vec{k}',\vec{k}\neq\vec{k}'}^{\mathcal{S}_1} J_{\vec{k},\vec{k}'}\sigma_{\vec{k}} \sigma_{\vec{k}'}$. The ground state corresponding to the optimal spin-configuration is obtained from~\equref{eq:HK_degenerate_gs}. It is worth mentioning that any many-body state $\ket{\{ \sigma_{\vec{k}} \}}$ breaks $\mathcal{PT}$ symmetry. Additionally, since the HK ground state is characterized by a fixed number of electrons at each momentum, the translation invariance is conserved for all the states (also beyond the aforementioned MF states) within this subspace. This means that antiferromagnetism is not accessible by first order perturbation theory in the discussed setting.

\begin{figure}[t] 
    \centering
    \includegraphics[width=\columnwidth]{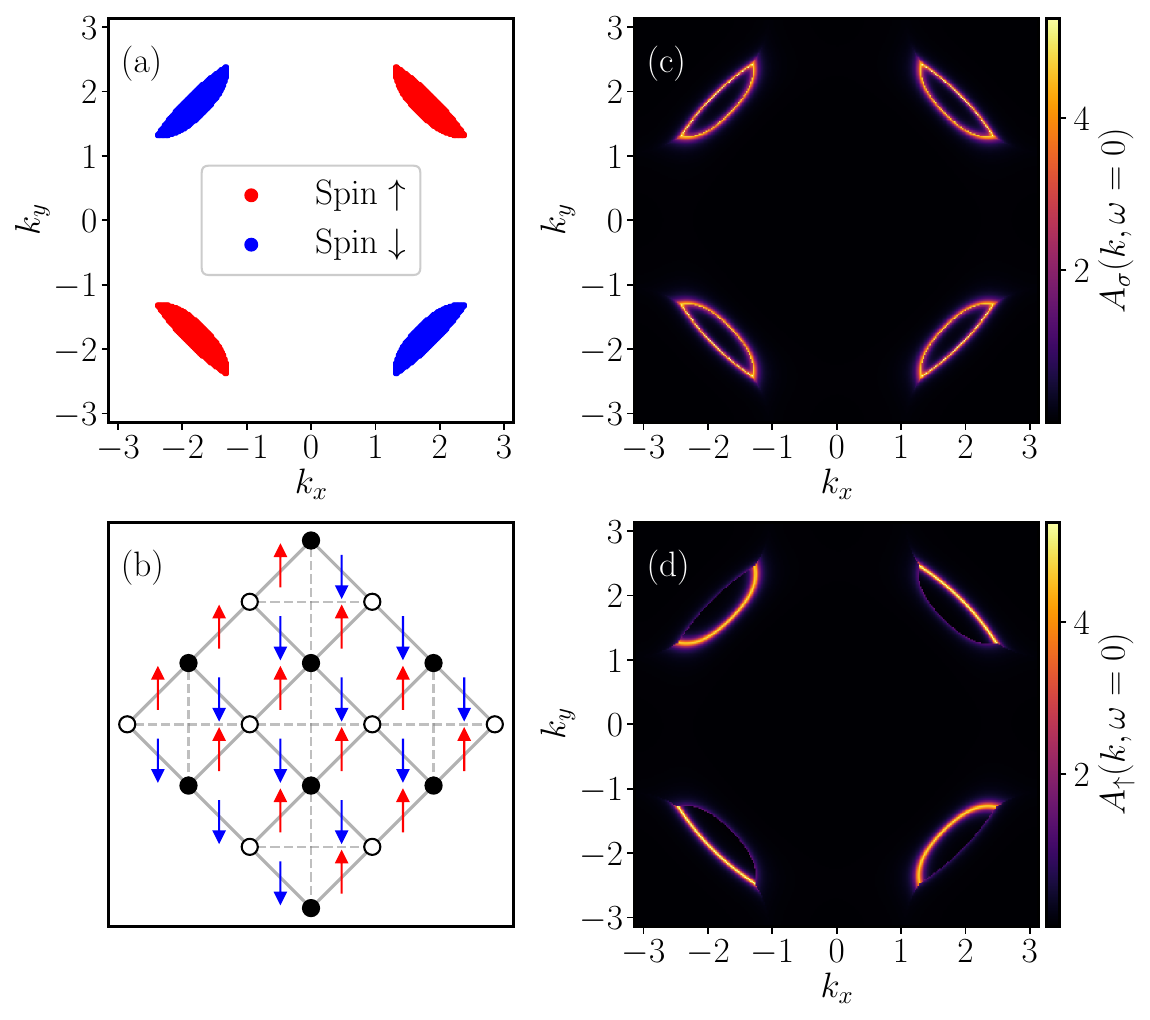}
    \caption{\textbf{(a)} $d$-wave spin configuration stabilized by $\Delta H$ in \equref{eq:deltaH} in the $l_d$ regime, denoted by $l_d$-$d$. The associated altermagnetic real-space bond order is displayed in \textbf{(b)}. \textbf{(c)}~Spectral function $A_\sigma (\vec{k},\omega=0)$ for the bare HK Hamiltonian in the $l_d$ regime, which obeys $A_\uparrow = A_\downarrow = A_\sigma$.
    \textbf{(d)} Spin-up spectral function $A_\uparrow (\vec{k},\omega=0)$ for the perturbed HK in the $l_d$-$d$ phase of (a); $A_\downarrow (\vec{k},0)$ is be obtained from $A_{\uparrow}(\vec{k},0)$ by $\pi/2$ rotation around the center of the BZ. For all figures, we chose $U=-1.0$, $U^\prime = 4.0$, $\mu=-1.0$, and $t_2=0.5$.}
    \label{fig:DWave}
\end{figure}

We first specifically consider the $l_d$ regime of the ground state occupation. For $(U, U') = (-1, 4)$, the minimization of the Ising model yields the spin-configuration presented in~\figref{fig:DWave}a. This alternating magnetic texture breaks time-reversal ($\mathcal{T}$) but preserves inversion ($\mathcal{P}$) symmetry, which makes it an altermagnet \cite{ReviewUnconventionalMagnetism}; specifically of $d$-wave form. Interestingly, this does not translate into the conventional on-site altermagnetic order, as $\braket{c^\dagger_{\vec{R}+\vec{a}_l} \sigma_{z} c^{\pd}_{\vec{R}+\vec{a}_{l}}} = 0$, but rather leads to the spin-bond texture with real-valued non-vanishing spins on the nearest-neighbor bonds, $\braket{c^\dagger_{\vec{R}} \sigma_{z} c^{\pd}_{\vec{R} + \vec{\delta}_j}} \neq 0$, forming the alternating magnetic pattern shown in \figref{fig:DWave}b. 

\begin{figure}[t] 
    \centering
    \includegraphics[width=\columnwidth]{Figures/Fig3_upd.pdf}
    \caption{\textbf{(a)} $A_\uparrow (\vec{k},\omega=0)$ for the perturbed HK              Hamiltonian in the $l_d$-$p$ phase, i.e., $p$-wave magnetism in the $l_d$ regime ($U=-1.0$, $U^\prime =9.0$). \textbf{(b)} Same quantity for the $l_d$-$\mathrm{FM}$ state, with interactions $U=-1.75$, $U^\prime =20.0$. \textbf{(c)} Phase diagram of the different magnetic phases, defined by the combination of the type of $\mathcal{S}_1$ region for the underlying unperturbed HK model ($l_d$, $l_v$, ring, and disk) and the form of magnetic order (FM, $d$, and $p$) obtained through energetic minimization. Throughout, we chose $\mu=-1.0$, $t_2 = 0.5$.
                }
    \label{fig:SpinSplitting}
\end{figure}

To demonstrate how the discussed symmetry breaking affects the spectral properties of the system, we construct the MF version of the Hamiltonian~\label{eq:hubbard_perturbation} (via Hartree-Fock decomposition) based on the correlations and order parameters of the MF ground state $\ket{\{ \sigma_{\vec{k}} \}^{\min}}$ obtained by minimizing the Ising model. The spin-splitting magnetic part of the MF Hamiltonian can be written as $\Delta H_{\mathrm{MF}}^{\mathrm{m}}[\{ \sigma_{\vec{k}} \}^{\min}] = \sum_{\vec{k},l,l'} \bigl( S_{\vec{k},z}^{l,l'} B_{\vec{k}}^{l',l}[\{ \sigma_{\vec{k}} \}^{\min}] + \text{H.c.} \bigr)$, where $S_{\vec{k}}^{l,l'} = \frac{1}{2} c^\dagger_{\vec{k},l} \sigma_z c_{\vec{k},l'}^{\pd}$, and $B_{\vec{k}}^{l',l}$ is the effective (alter)magnetic order parameter \cite{SIref}. Taking only this term into account (and neglecting additional terms that simply renormalize the bands), we compute the spin-resolved spectral function $A_{\sigma}(\vec{k},\omega)$ corresponding to $H_{\mathrm{HK}} + \Delta H_{\mathrm{MF}}^{\mathrm{m}}$; it is shown for $\sigma=\uparrow$ in~\figref{fig:DWave}d, and contrasted with the spectral function of the unperturbed HK model displayed in~\figref{fig:DWave}c. We clearly see that the long-range order significantly modifies $A_{\sigma}(\vec{k},0)$ by making it anisotropic in each pocket, following the $d$-wave symmetry of the altermagnet: the outer (inner) boundaries of the $\mathcal{S}_1$ lobes occupied by electrons with spin $\sigma$ (spin $-\sigma$) become more pronounced in $A_{\sigma}(\vec{k},0)$.

Remarkably, $l_d$ lobes can also accommodate other phases: a ferromagnet (FM) [$\sigma_{\vec{k}}$ independent of $\vec{k}$ in \equref{eq:HK_degenerate_gs}] and a $p$-wave magnet, where the spin polarization is opposite at opposite momenta, i.e., $\sigma_{\vec{k}} = -\sigma_{-\vec{k}}$. The $p$-wave state conserves $\mathcal{T}$, but breaks both $\mathcal{P}$ and $\mathcal{PT}$ symmetries.  This results in purely imaginary spin-bonds, $\mathrm{Im} \braket{c^\dagger_{\vec{R}} \sigma_{z} c^{\pd}_{\vec{R}+\vec{\delta}_j}} \neq 0$, $\mathrm{Re}\braket{c^\dagger_{\vec{R}} \sigma_{z} c^{\pd}_{\vec{R}+\vec{\delta}_j}} = 0$ with vanishing on-site magnetization. Therefore, we can associate this order with emerging spin-orbit coupling and spontaneous generation of spin-currents. For the FM, the uniform spin polarization of the lobes manifests as conventional on-site order with $\braket{c^{\dagger}_{\vec{R}+\vec{a}_l} \sigma_z c^{\pd}_{\vec{R}+\vec{a}_l}} \neq 0$. This state even becomes exact (beyond MF) when $J_{\vec{k},\vec{k}'} > 0, \forall \vec{k},\vec{k}'\in\mathcal{S}_1$ (e.g., when there is only on-site repulsive interaction) \footnote{To be precise, our perturbative approach requires first setting $\lambda\rightarrow 0$ before taking the thermodynamic limit so that the finite size gap close to the boundary of $\mathcal{S}_1$ is large compared to $\lambda V^{(')}$. However, even in the thermodynamic limit, the fraction of $\mathcal{S}_1$ where our approach is uncontrolled vanishes for $\lambda \rightarrow 0$ and so we expect no qualitative changes in the ordering tendencies reported in Figs.~\ref{fig:DWave} and \ref{fig:SpinSplitting}.}. We present the corresponding spin-resolved spectral functions in~\figref{fig:SpinSplitting}(a-b), respectively.

Finally, \figref{fig:SpinSplitting}c shows the phase diagram capturing the different combinations of ground state occupation and magnetic phase. Since the ground state depends sensitively on the behavior of the coupling constant $J_{\vec{k},\vec{k}'}$ within $\mathcal{S}_1$, the phase diagram is susceptible to changes in $\mu$ and $h_{\vec{k}}$ in general. For instance, if $\mathcal{S}_1 \in \{\vec{k} : |k_i| < \pi/2\}$, the coupling constants $J_{\vec{k},\vec{k}'}$ become positive for arbitrary $U,U'>0$, and therefore, only FM can appear regardless of the form of $\mathcal{S}_1$. We also find that the standard single-orbital HK model can host all the above mentioned MF states \cite{SIref}. The corresponding phase diagram is, however, less rich, as there is less control of the shape and position of the singly occupied region, which can only be either a ring or a disk.

\vspace{1em}

\textit{Non-magnetic regime}---We have seen that conventional interactions~\equref{eq:deltaH} projected onto the degenerate HK ground state map to effective spin models. 
Being dense and generally frustrated, these models lead to complex phase diagrams but require approximate techniques for diagonalization. In analogy to extended HK models with density-density couplings across a finite numbers of momenta \cite{PhysRevLett.133.166501,TwistedHubbard,2026arXiv260406588M}, we replace \equref{EmergentSpinModel} by 
\begin{align}
    \delta H = - \sum_{C} \sum_{\vec{k},\vec{k}'\in C}  \sum_{l_1,l_2,l_1',l_2'} \tilde{J}_{\vec{k},\vec{k}'}^{l_1l_2l_1'l_2'}\, \vec{S}_{\vec{k}}^{l_1,l_2} \cdot \vec{S}_{\vec{k}'}^{l_1',l_2'}, \label{eq:second_perturbation}
\end{align}
where the summation $\sum_{C}$ goes over non-overlapping finite-size clusters $C$ of coupled momenta. For small cluster sizes $|C|$, the combined model $H_{\mathrm{HK}} + \delta H$ is still readily diagonalized exactly. As we will see next, such additional interactions can substantially modify the ground-state and spin-spin correlations.

\begin{figure}[t] 
    \centering
    \includegraphics[width=\columnwidth]{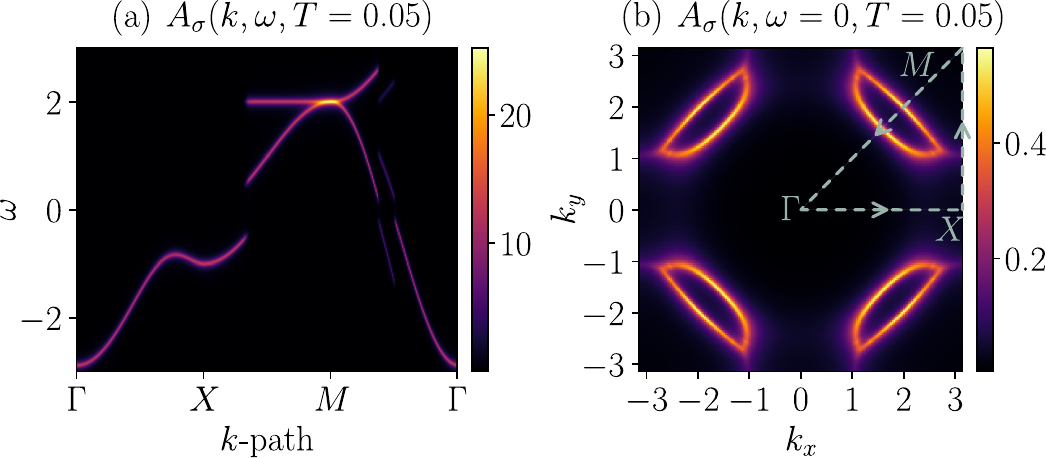}
    \caption{\textbf{(a)}  Spectral function $A_\sigma (\vec{k},\omega)$ of $H_{\text{HK}}+\delta H_{2S}$ along high symmetry paths in $\vec{k}$ space and \textbf{(b)} its momentum dependence at $\omega=0$, i.e., $A_\sigma (\vec{k},\omega=0)$. Note that the perturbation $\delta H_{2S}$ leads to the emergence of a gap but does not break spin-rotation symmetry, $A_\uparrow=A_\downarrow=A_\sigma$.
                Both spectral functions are computed at finite temperature $T=0.05$ and coupling $J~=~-1$. The parameters of the bare HK model are chosen to match those of the spectral function in \figref{fig:DWave}c.
                }
    \label{fig:Spectrals}
\end{figure}

We first focus on the specific example where each $\vec{k}$ is only coupled with $-\vec{k}$ and $\delta H_{2S} = -J \sum_{\vec{k},l} \vec{S}_{\vec{k}}^{l,l} \cdot \vec{S}_{-\vec{k}}^{l,l}$. It is straightforward to see that the ground-state degeneracy of $H_{\mathrm{HK}}$ is fully lifted already in first order perturbation theory in $J<0$; it leads to a non-magnetic
ground state consisting of decoupled singlet states formed from electrons with $\vec{k}$ and $-\vec{k}$ within $\mathcal{S}_1$,
\begin{align}
    \ket{\mathrm{GS}} = \prod_{\vec{k}\in\mathcal{S}_1,k_x>0} \frac{1}{\sqrt{2}} \left[ f^\dagger_{\vec{k},\uparrow} f^{\dagger}_{-\vec{k},\downarrow} - f^\dagger_{\vec{k},\downarrow} f^{\dagger}_{-\vec{k},\uparrow} \right] \ket{\mathcal{S}_2}. \label{eq:k_minus_k_perturbation}
\end{align}
Since $|C|$ is only two, it is straightforward to numerically compute the spectral properties of $H_{\mathrm{HK}} + \delta H_{2S}$ exactly, i.e., beyond perturbation theory. As can be seen in \figref{fig:Spectrals}a, where we took the same parameters for $H_{\mathrm{HK}}$ as in \figref{fig:DWave}c, the low-energy spectral weight is suppressed between $\Gamma$ and $M$. While this suppression comes from the gap induced by $J$, it still allows for peaks in $A_\sigma(\vec{k},\omega=0)$ for finite temperature $T$ (and for finite $\omega$). As shown in \figref{fig:Spectrals}b, these peaks look similar to those of $\sum_\sigma A_\sigma(\vec{k},\omega)$ of the related ordered unconventional magnets stabilized by \equref{EmergentSpinModel}.
We have therefore constructed an exactly solvable scenario with a unique, non-magnetic ground state and spectral properties reminiscent of unconventional magnets. This is phenomenologically reminiscent of the ``fractionalized itinerant altermagnets'' envisioned in \refcite{PhysRevResearch.7.023152}, where it was discussed on the level of an effective parton theory. We note that the theory \cite{PNASPseudogap} underlying \cite{PhysRevResearch.7.023152} also has a gap around $\omega =0$ for $T\rightarrow 0$.

Finally, we address the static spin-structure factor, $S(\vec{q}) = N^{-1} \braket{ \vec{M}_{\vec{q}} \cdot \vec{M}_{-\vec{q}} }$, $\vec{M}_{\vec{q}} = \frac{1}{2} \sum_{\vec{k},l} c^\dagger_{\vec{k}+\vec{q},l} \boldsymbol{\sigma} c^{\pd}_{\vec{k},l}$, which can be compactly expressed for the general form \equref{eq:second_perturbation} of $\delta H$, using that only spins in the same cluster are correlated. For instance, for $\vec{q}=0$, one finds  
\begin{align}
    S(\vec{q}=\vec{0}) = \frac{1}{N} \sum_{C} \Braket{ \biggl( \sum_{\vec{k} \in C} \vec{S}_{\vec{k}} \biggr)^2 },
\end{align}
where $\vec{S}_{\vec{k}} = \sum_{l}\vec{S}_{\vec{k}}^{l,l}$. We can thus immediately see that, e.g., $P \delta H_{2S} P$ fully suppresses $S(\vec{q}=\vec{0})$ due to the singlet formation in \equref{eq:k_minus_k_perturbation}. For general $\vec{q}$, $P\delta H P$ yields a contribution if there are $\vec{k}\in \mathcal{S}_1$ with $\vec{k}+\vec{q} \in \mathcal{S}_1 \cap C_{\vec{k}}$ where $C_{\vec{k}}$ is the cluster with $\vec{k} \in C_{\vec{k}}$. 

For our example $\delta H_{2S}$, this is the case for $\vec{k}$ with $\vec{k} = \vec{q}/2$, which again lowers $S(\vec{q})$ due to singlet formation. This effect, however, vanishes in the thermodynamic limit as only finitely many momenta contribute at given $\vec{q} \neq 0$. This is different for interactions like $\delta H_{\vec{Q}} = -J \sum_{\vec{k},l} \vec{S}_{\vec{k}}^{l,l}\cdot\vec{S}_{\vec{k}+\vec{Q}}^{l,l}$ that generate clusters $C_{\vec{k}} = \{ \vec{k} + m\vec{Q} | m\in\mathds{Z}\}$ (modulo reciprocal lattice vectors), which for $J<0$ can coherently suppress $S( m \vec{Q})$ at integer $m$.

\vspace{1em}
\textit{Conclusion}---We have shown how HK-inspired models can be used as a transparent starting point to capture the formation of and interaction effects in itinerant unconventional magnets and related non-magnetic states. We believe that our work can further provide the starting for many interesting follow-up works, e.g., on the interplay of pairing with altermagnetism in the strong-coupling regime.

\begin{acknowledgments}
M.S.S. thanks B.~Putzer and J.~Sobral for discussions. D.S.~and M.S.S.~acknowledge funding by the European Union (ERC-2021-STG, Project 101040651---SuperCorr). Views and opinions expressed are however those of the authors only and do not necessarily reflect those of the European Union or the European Research Council Executive Agency. Neither the European Union nor the granting authority can be held responsible for them.
\end{acknowledgments}

\bibliography{draft_Refs}

\onecolumngrid

\begin{appendix}

\section{Solving the altermagnetic HK model}
\subsection{Finding the spectrum of the altermagnetic HK model}
Let us consider the HK Hamiltonian of our system
\begin{equation}
    H_{\text{HK}} =    \sum_{\boldsymbol{k}} \sum_{l,l^\prime,\sigma} c^\dagger_{\boldsymbol{k},l,\sigma} h_{\vec{k}}^{l,l'} c_{\boldsymbol{k},l^\prime,\sigma} -
                \mu \sum_{\boldsymbol{k},l,\sigma} n_{\boldsymbol{k},l,\sigma}
                + U \sum_{\boldsymbol{k},l} n_{\boldsymbol{k},l,\uparrow} n_{\boldsymbol{k},l,\downarrow}
                + U^\prime \sum_{\boldsymbol{k}} \sum_{\sigma, \sigma^\prime} n_{\boldsymbol{k},A,\sigma} n_{\boldsymbol{k},B,\sigma^\prime}~. \label{eq:HK_checkerboard_si}
\end{equation}
We obtain the tight binding matrix $h_{\vec{k}}^{l,l'} $ by considering nearest neighbor hopping ($t_1$) between atoms A and B as well as direction dependent next-nearest neighbor hopping ($t_2$) between atoms of the same sublattice. Atoms of the sublattice A can hop with amplitude $t_2$ in the $\boldsymbol{e}_x$ direction whereas atoms of the sublattice B hop in the $\boldsymbol{e}_y$ direction. 
The tight binding matrix is then given as 
\begin{equation}
    h_{\vec{k}}^{l,l'}  = \begin{pmatrix}
                                            p(k_x) & f(k_x,k_y) \\
                                            f(k_x,k_y) & g(k_y)
                                        \end{pmatrix}_{(l,l^\prime)} 
                                     = \begin{pmatrix}
                                            -2t_2 \cos \left(k_x\right) & -4t_1 \cos \left(\frac{k_x}{2}\right) \cos \left(\frac{k_y}{2}\right) \\
                                            -4t_1 \cos \left(\frac{k_x}{2}\right) \cos \left(\frac{k_y}{2}\right) & -2t_2 \cos \left(k_y\right)
                                        \end{pmatrix}_{(l,l^\prime)} 
\end{equation}
where the Fourier transformation is defined as
\begin{equation}
    c^\dagger_{\boldsymbol{R}+\vec{a}_l,\sigma} = \frac{1}{N} \sum_{\boldsymbol{k}} e^{-i \boldsymbol{k} \cdot (\vec{R}+\vec{a}_l)} c^\dagger_{\boldsymbol{k},l,\sigma}~~~~\mathrm{and}~~~~
    c_{\boldsymbol{R}+\vec{a}_l,\sigma} = \frac{1}{N} \sum_{\boldsymbol{k}} e^{i \boldsymbol{k} \cdot (\vec{R}+\vec{a}_l)} c_{\boldsymbol{k},l,\sigma}. \label{eq:fourier_transform_convention_si}
\end{equation} 
Here $\vec{R}$ are Bravais square lattice vectors, and $\vec{a}_A = 0$, $\vec{a}_B = (1,1)^T/2$ are the positions of the sublattices' atoms in the unit cell.

Using that $H_{\text{HK}} = \sum_{\boldsymbol{k}} H_{\boldsymbol{k}}$ is block-diagonal in momentum $\boldsymbol{k}$, we can diagonalize it independently for every momentum. This equates to diagonalizing a $16\times 16$ ($16=2^4$, where 4 is the number of modes: two orbital and two spin) matrix at every momentum. 
With $\left[H_{\boldsymbol{k}},N_{\boldsymbol{k}}\right] = 0$ this can be further diagonalized in total particle number $N_{\boldsymbol{k}}$ blocks ($N_{\boldsymbol{k}}=\sum_{l,\sigma} n_{\boldsymbol{k},l,\sigma}$), which allows us to find analytic expressions for the eigenvalues of $H_{\boldsymbol{k}}$.

For this, let us define the new quantities
\begin{align}
    \Omega(k_x,k_y):= \frac{p-g}{2f}, &\hspace{15pt} Q(k_x,k_y) := -\frac{\Delta U^2}{9}- \frac{4}{3} f^2 (1+\Omega^2), \\
    R(k_x,k_y) := - \frac{\Delta U^3}{27} - \frac{2}{3} f^2 \Delta U (1-2\Omega^2), &\hspace{15pt} \theta(k_x,k_y) := \arccos \left(\frac{R}{\sqrt{-Q^3}}\right)~.
\end{align}
Then the eigenenergies in the different particle-number blocks are given as: \\
\begin{itemize}

\item \textbf{$N=0$}
\begin{equation}
E_0 = 0
\end{equation}

\item \textbf{$N=1$}
\begin{align}
E_1^1 &= \frac{p+g}{2} - \sqrt{f^2 (1+\Omega^2)} - \mu \\
E_1^2 &= \frac{p+g}{2} - \sqrt{f^2 (1+\Omega^2)} - \mu \\
E_1^3 &= \frac{p+g}{2} + \sqrt{f^2 (1+\Omega^2)} - \mu \\
E_1^4 &= \frac{p+g}{2} + \sqrt{f^2 (1+\Omega^2)} - \mu
\end{align}

\item \textbf{$N=2$}
\begin{equation}
E_2^0(k_x,k_y) = p + g + U' - 2\mu
\end{equation}

\begin{align}
E_2^1 &= E_2^0 \\
E_2^2 &= E_2^0 \\
E_2^3 &= 2\sqrt{-Q}\cos\!\left(\frac{\theta}{3}\right) 
        + \frac{2\Delta U}{3} + E_2^0 \\
E_2^4 &= 2\sqrt{-Q}\cos\!\left(\frac{\theta + 2\pi}{3}\right) 
        + \frac{2\Delta U}{3} + E_2^0 \\
E_2^5 &= E_2^0 \\
E_2^6 &= 2\sqrt{-Q}\cos\!\left(\frac{\theta + 4\pi}{3}\right) 
        + \frac{2\Delta U}{3} + E_2^0
\end{align}

\item \textbf{$N=3$}
\begin{equation}
E_3^0(k_x,k_y) = p + g + U + 2U' - 3\mu
\end{equation}

\begin{align}
E_3^1 &= \frac{p+g}{2} - \sqrt{f^2 (1+\Omega^2)} + E_3^0 \\
E_3^2 &= \frac{p+g}{2} - \sqrt{f^2 (1+\Omega^2)} + E_3^0 \\
E_3^3 &= \frac{p+g}{2} + \sqrt{f^2 (1+\Omega^2)} + E_3^0 \\
E_3^4 &= \frac{p+g}{2} + \sqrt{f^2 (1+\Omega^2)} + E_3^0
\end{align}

\item \textbf{$N=4$}
\begin{equation}
E_4 = 2p + 2g + 2U + 4U' - 4\mu
\end{equation}

\end{itemize}

\subsection{Band basis of the checker board lattice}

Up to this point, the creation and annihilation operators act on the different sublattices $l\in \{A,B\}$. They can be transformed with a unitary transformation $U$ into the band basis of the Bloch Hamiltonian $h_{\boldsymbol{k}}$ where they act on the two bands $\alpha \in\{\pm\}$,
\begin{equation}
    U^T(\boldsymbol{k})  h_{\boldsymbol{k}} U(\boldsymbol{k})  = \begin{pmatrix}
\lambda_- (\boldsymbol{k}) & 0 \\
0 & \lambda_+(\boldsymbol{k}) 
\end{pmatrix}
\end{equation}
with the eigenvalues $\lambda_\pm$ of the Bloch Hamiltonian $h_{\boldsymbol{k}} $.
The eigenvalues are given as
\begin{equation}
    \lambda_\pm = \frac{p+g}{2} \pm \sqrt{f^2 + \left(\frac{g-p}{2}\right)^2}
\end{equation}
and with this, the transformation matrix can be written as 
\begin{equation}
    U_{l,\alpha}  = \begin{pmatrix}
U_{A,-} & U_{A,+} \\
U_{B,-} & U_{B,+}
\end{pmatrix} = \begin{pmatrix}
\frac{f}{N_-} & \frac{f}{N_+} \\
\frac{\lambda_- - p}{N_-} & \frac{\lambda_+ - p}{N_-}
\end{pmatrix} \label{eq:hk_eigenstates_si}
\end{equation}
where the normalizations are given as $N_\pm~=~\sqrt{f^2 + (\lambda_\pm-p)^2}$. It can be clearly seen that the transformation matrix elements depend on the momentum $\vec{k}$.
With the transformation matrix, we can then express the orbital basis operators $(c^\dagger / c)$ in terms of the new band operators $(f^\dagger / f)$. With this we can write the ground state as in Eq. 2 of the main text. The mapping is given as
\begin{equation}
    c^{(\dagger)}_{\vec{k},l,\sigma} = \sum_\alpha U_{l,\alpha} (\vec{k}) f^{(\dagger)}_{\vec{k},\alpha,\sigma}.
\end{equation}
Note that there is no need for complex conjugation as the transformation matrix $U$ is real.

\section{Derivation of the Heisenberg model}

In this section, we show that the Coulomb-like interaction projected on the degenerate ground state of the HK model can be mapped onto the spin-1/2 quantum Heisenberg model.

\subsection{One-site HK model}

We first consider the one-site HK model,
\begin{align}
    H_{\mathrm{HK}} = \sum_{\vec{\vec{k}},\sigma} \xi_{\vec{k}} \hat{n}_{\vec{k},\sigma} + \sum_{\vec{k}} \hat{n}_{\vec{k},\uparrow} \hat{n}_{\vec{k},\downarrow}.
\end{align}
The degenerate ground states of this model are given by
\begin{align}
    \ket{\{ \sigma_{\vec{k}} \}} = \prod_{\vec{k}\in\mathcal{S}_1} c^\dagger_{\vec{k},\sigma_{\vec{k}}} \prod_{\vec{k}' \in \mathcal{S}_{2}} c^\dagger_{\vec{k},\uparrow} c^\dagger_{\vec{k},\downarrow} \ket{0},
\end{align}
where $\mathcal{S}_{1,2}$ denote the singly and doubly occupied regions of the BZ, respectively. 

Let us consider the following Coulomb-like interaction
\begin{align}
    \Delta H = \frac{1}{N} \sum_{\vec{k},\vec{k}',\vec{q},\sigma,\sigma'} \mathcal{V}_{\vec{q}} : c^\dagger_{\vec{k}+\vec{q},\sigma} c^{\pd}_{\vec{k},\sigma} c^\dagger_{\vec{k}'-\vec{q},\sigma'} c^{\pd}_{\vec{k}',\sigma'}:, \label{eq:H_int_one_site_si}
\end{align}
where $:\hat{A}:$ denotes normal ordering. We want to project the Hamiltonian~\eqref{eq:H_int_one_site_si} onto the degenerate HK ground state. Let $P$ be the corresponding projector, $P = \sum_{\{\sigma_{\vec{k}}\}} \ket{\{ \sigma_{\vec{k}} \}} \bra{\{ \sigma_{\vec{k}} \}}$. Since this subspace is characterized by a fixed number of electrons at each momentum $\vec{k}$, the projected Hamiltonian conserves the number of particles $\hat{n}_{\vec{k}} = \sum_{\sigma}\hat{n}_{\vec{k},\sigma}$. Therefore, only the terms with $\vec{q} = 0$ and $\vec{q} = \vec{k}'-\vec{k} \neq 0$ in~\eqref{eq:H_int_one_site_si} survive after the projection, $P \Delta H P$. Let us study these two contributions separately, 
\begin{align}
    P \Delta H P = \Delta \tilde{H}_{\vec{q}=0} + \Delta \tilde{H}_{\vec{q}=\vec{k}'-\vec{k}}, \label{eq:projected_h_survived_terms}
\end{align}
where 
\begin{align}
   \Delta \tilde{H}_{\vec{q}=0} = P \frac{ \mathcal{V}_{0}}{N}  \sum_{\vec{k},\vec{k}',\sigma,\sigma'} : c^\dagger_{\vec{k},\sigma} c^{\pd}_{\vec{k},\sigma} c^\dagger_{\vec{k}',\sigma'} c^{\pd}_{\vec{k}',\sigma'} : P = \frac{\mathcal{V}_0}{N} \sum_{\vec{k},\vec{k}'} P : \hat{n}_{\vec{k}} \hat{n}_{\vec{k}'} : P = \mathrm{const.} P,
\end{align}
and
\begin{align}
    \Delta \tilde{H}_{\vec{q}=\vec{k}'-\vec{k}} &= P \frac{1}{N}
    \sum_{\vec{k},\vec{k}',\vec{k}\neq\vec{k}',\sigma,\sigma'} \mathcal{V}_{\vec{k}'-\vec{k}} : c^\dagger_{\vec{k}',\sigma} c^{\pd}_{\vec{k},\sigma} c^\dagger_{\vec{k},\sigma'} c^{\pd}_{\vec{k}',\sigma'} : P \nonumber \\
    &= - P \frac{1}{N} \sum_{\vec{k},\vec{k}',\vec{k}\neq\vec{k}',\sigma,\sigma'} \mathcal{V}_{\vec{k}'-\vec{k}} c^\dagger_{\vec{k}',\sigma} c^{\pd}_{\vec{k}',\sigma'} c^\dagger_{\vec{k},\sigma'} c^{\pd}_{\vec{k},\sigma} P
\end{align}
We can rewrite the remaining non-trivial term in the following way,
\begin{align}
    -P\sum_{\vec{k},\vec{k}',\vec{k}\neq\vec{k}',\sigma,\sigma'} \mathcal{V}_{\vec{k}'-\vec{k}} c^\dagger_{\vec{k}',\sigma} c^{\pd}_{\vec{k}',\sigma'} c^\dagger_{\vec{k},\sigma'} c^{\pd}_{\vec{k},\sigma} P &= -
     \sum_{\vec{k},\vec{k}',\vec{k}\neq\vec{k}'} 2 \mathcal{V}_{\vec{k}'-\vec{k}} P \left[ \frac{1}{4} \hat{n}_{\vec{k}} \hat{n}_{\vec{k}'} + \vec{S}_{\vec{k}} \cdot \vec{S}_{\vec{k}'} \right] P \nonumber \\
     &= \mathrm{const.} P - \sum_{\vec{k},\vec{k}',\vec{k}\neq\vec{k}'} 2 \mathcal{V}_{\vec{k}'-\vec{k}} P \vec{S}_{\vec{k}} \cdot \vec{S}_{\vec{k}'}  P,
\end{align}
where $\vec{S}_{\vec{k}} = \frac{1}{2} c^\dagger_{\vec{k}} \boldsymbol{\sigma} c^{\pd}_{\vec{k}}$. The second term is nonzero only when $\vec{k},\vec{k}' \in\mathcal{S}_1$, where it can be directly mapped onto the spin-1/2 quantum Heisenberg model. Thus,
\begin{align}
    P \Delta H P \mapsto - \frac{1}{2N} \sum_{\vec{k},\vec{k}' \in \mathcal{S}_1,\vec{k}\neq\vec{k}'} \mathcal{V}_{\vec{k}-\vec{k}'} \boldsymbol{\sigma}_{\vec{k}} \cdot \boldsymbol{\sigma}_{\vec{k}'} + \mathrm{const}.
\end{align}

\subsection{Multiorbital HK model}

We now generalize the obtained mapping to the multiorbital HK models,
\begin{align}
    H_{\mathrm{HK}} = \sum_{\vec{\vec{k}},\sigma,\alpha,\beta} c^\dagger_{\vec{k},\alpha,\sigma} h_{\vec{k}}^{\alpha\beta} c^{\pd}_{\vec{k},\beta,\sigma} + \sum_{\vec{k},\alpha,\beta,\sigma,\sigma'} \hat{n}_{\vec{k},\alpha,\sigma} U^{\alpha\beta} \hat{n}_{\vec{k},\beta,\sigma'}, \label{eq:multiorbital_HK}
\end{align}
where $h_{\vec{k}}^{\alpha\beta}$ is a one-particle Hamiltonian in the orbital basis, and $U^{\alpha\beta}$ is an interaction matrix. We limit our analysis to the case when the ground states have a non-degenerate doubly-occupied region $\mathcal{S}_2$ and a spin-degenerate singly occupied region $\mathcal{S}_1$. The degenerate ground states are then given by
\begin{align}
    \ket{\{ \sigma_{\vec{k}} \}} = \prod_{\vec{k}\in\mathcal{S}_1} f^\dagger_{\vec{k},1,\sigma_{\vec{k}}} \ket{\mathcal{S}_2}, \label{NewStatesToProjectOn}
\end{align}
where $\ket{\mathcal{S}_2}$ is the doubly-occupied part of the ground state, and $f^\dagger_{\vec{k},1,\sigma_{\vec{k}}}$ is the creation operator of the lowest band of the Bloch Hamiltonian $h_{\vec{k}}^{\alpha\beta}$. Note that we do not need to know the exact structure of the ground state wavefunction in the doubly occupied region.

We consider the multiorbital Hubbard model with the Coulomb-like interaction,
\begin{align}
    \Delta H = \frac{1}{N} \sum_{\vec{k},\vec{k}',\vec{q},\sigma,\sigma'} \mathcal{V}_{\vec{q}}^{\alpha\beta} : c^\dagger_{\vec{k}+\vec{q},\alpha,\sigma} c^{\pd}_{\vec{k},\alpha,\sigma} c^\dagger_{\vec{k}'-\vec{q},\beta,\sigma'} c^{\pd}_{\vec{k}',\beta,\sigma'}:, \label{eq:H_int}
\end{align}
where the summation over orbitals $\alpha,\beta$ is assumed. Due to the conservation of $\hat{n}_{\vec{k}}$, only two terms survive the projection onto the degenerate ground states in \equref{NewStatesToProjectOn}. Let us analyze the first term
\begin{align}
   \Delta \tilde{H}_{\vec{q}=0} = \frac{ \mathcal{V}_{0}^{\alpha\beta}}{N}  \sum_{\vec{k},\vec{k}'} P : \hat{n}_{\vec{k},\alpha} \hat{n}_{\vec{k}',\beta} : P,
\end{align}
where $\hat{n}_{\vec{k},\alpha} = \sum_{\sigma} \hat{n}_{\vec{k},\alpha,\sigma}$. In the subspace of interest, different $\vec{k}$ momenta lying in $\mathcal{S}_2$ are uncorrelated; hence, for these momenta,
\begin{align}
    P : \hat{n}_{\vec{k},\alpha} \hat{n}_{\vec{k}',\beta} : P = \delta_{\vec{k},\vec{k}'} \left[ \braket{  \hat{n}_{\vec{k},\alpha} \hat{n}_{\vec{k},\beta}} - \delta_{\alpha,\beta} \braket{\hat{n}_{\vec{k},\alpha}}\right] P + (1 - \delta_{\vec{k},\vec{k}'}) \braket{\hat{n}_{\vec{k},\alpha}} \braket{\hat{n}_{\vec{k}',\beta}} P,\quad \vec{k},\vec{k}' \in \mathcal{S}_2. \label{eq:n_term_decomposition}
\end{align}
When one or both of the momenta lie in the singly occupied region, the projection requires more accurate treatment. Let $f_{\vec{k},i}$ be band operators, with $f_{\vec{k},1}$ corresponding to the lowest one-particle band. In the following calculations, we denote the wavefunctions of this band by $\phi_{\vec{k},\alpha}$. For $\vec{k} \in \mathcal{S}_1$, $ f_{\vec{k},i} P \propto \delta_{i,1}, P f^\dagger_{\vec{k},i} \propto \delta_{i,1}$, therefore,
\begin{align}
    P : \hat{n}_{\vec{k},\alpha} \hat{n}_{\vec{k}',\beta} : P =  | \phi_{\vec{k},\alpha}|^2 \braket{\hat{n}_{\vec{k}',\beta}} P \hat{n}_{\vec{k}, f_1} P  = | \phi_{\vec{k},\alpha}|^2 \braket{\hat{n}_{\vec{k}',\beta}} P,\quad \vec{k}'\in\mathcal{S}_2,\vec{k}\in\mathcal{S}_1,
\end{align}
and
\begin{align}
    P : \hat{n}_{\vec{k},\alpha} \hat{n}_{\vec{k}',\beta} : P &= | \phi_{\vec{k},\alpha}|^2 |\phi_{\vec{k}',\beta}|^2   P \hat{n}_{\vec{k},f_1} \hat{n}_{\vec{k}',f_1}  P - \delta_{\vec{k},\vec{k}'} \delta_{\alpha,\beta} |\phi_{\vec{k},\alpha}|^2 P \hat{n}_{\vec{k},f_1} P \nonumber \\
    &= | \phi_{\vec{k},\alpha}|^2 |\phi_{\vec{k}',\beta}|^2 P -  \delta_{\vec{k},\vec{k}'} \delta_{\alpha,\beta} | \phi_{\vec{k},\alpha}|^2 P,\quad \vec{k},\vec{k}'\in\mathcal{S}_1.
\end{align}
Thus, expression~\eqref{eq:n_term_decomposition} holds for an arbitrary combination of momenta. The second term that survives projection reads
\begin{align}
    \Delta \tilde{H}_{\vec{q}=\vec{k}'-\vec{k}} &= P \frac{1}{N}
    \sum_{\vec{k},\vec{k}',\vec{k}\neq\vec{k}',\sigma,\sigma'} \mathcal{V}_{\vec{k}'-\vec{k}}^{\alpha\beta} : c^\dagger_{\vec{k}',\alpha,\sigma} c^{\pd}_{\vec{k},\alpha,\sigma} c^\dagger_{\vec{k},\beta,\sigma'} c^{\pd}_{\vec{k}',\beta,\sigma'}:P.
\end{align}
We can rewrite it in the convenient way,
\begin{align}
    \sum_{\vec{k},\vec{k}',\vec{k}\neq \vec{k}} \sum_{\sigma,\sigma'} \mathcal{V}_{\vec{k}'-\vec{k}}^{\alpha\beta} :c^\dagger_{\vec{k}',\alpha,\sigma} c^{\pd}_{\vec{k},\alpha,\sigma} c^\dagger_{\vec{k},\beta,\sigma'} c^{\pd}_{\vec{k}',\beta,\sigma'} : &= - \sum_{\vec{k},\vec{k}',\vec{k}\neq \vec{k}} \sum_{\sigma,\sigma'} \mathcal{V}_{\vec{k}'-\vec{k}}^{\alpha\beta} c^\dagger_{\vec{k}',\alpha,\sigma}  c^{\pd}_{\vec{k}',\beta,\sigma'} c^{\dagger}_{\vec{k},\beta,\sigma'} c^{\pd}_{\vec{k},\alpha,\sigma}\nonumber \\
    &= - \sum_{\vec{k},\vec{k}',\vec{k}\neq \vec{k}} \mathcal{V}_{\vec{k}'-\vec{k}}^{\alpha\beta} \left[ \frac{1}{2} C^{\alpha\beta}_{\vec{k}'} C^{\beta\alpha}_{\vec{k}} + 2 \vec{S}^{\alpha\beta}_{\vec{k}'} \cdot \vec{S}^{\beta\alpha}_{\vec{k}}\right],
\end{align}
where $C^{\alpha\beta}_{\vec{k}} = \sum_{\sigma} c^{\dagger}_{\vec{k},\alpha,\sigma} c^{\pd}_{\vec{k},\beta,\sigma}, \vec{S}^{\alpha\beta}_{\vec{k}} = \frac{1}{2} c^\dagger_{\vec{k},\alpha} \boldsymbol{\sigma} c^{\pd}_{\vec{k},\beta}$
Following the same argument as before, for $\vec{k}\neq\vec{k}'$, we get
\begin{align}
    P C^{\alpha\beta}_{\vec{k}'} C^{\beta\alpha}_{\vec{k}} P = \braket{C^{\alpha\beta}_{\vec{k}'}} \braket{C^{\beta\alpha}_{\vec{k}}} P,
\end{align}
and
\begin{align}
    P \vec{S}^{\alpha\beta}_{\vec{k}'} \cdot \vec{S}^{\beta\alpha}_{\vec{k}} P =
    \begin{cases}
        \phi_{\vec{k}',\alpha}^* \phi_{\vec{k}',\beta} \phi_{\vec{k}',\beta}^* \phi_{\vec{k}',\alpha} P \vec{S}_{\vec{k}',f_1} \cdot \vec{S}_{\vec{k},f_1}P, \quad &\vec{k},\vec{k}'\in\mathcal{S}_1,\\
        0,\quad &\text{otherwise},
    \end{cases}
\end{align}
where $\vec{S}_{\vec{k},f_1} = \frac{1}{2} f^{\dagger}_{\vec{k},1} \boldsymbol{\sigma} f^{\pd}_{\vec{k},1}$.
Thus, we can again map the projected Hamiltonian onto the spin-$1/2$ quantum Heisenberg model,
\begin{align}
    P \Delta H P \mapsto \sum_{\vec{k},\vec{k}' \in \mathcal{S}_1,\vec{k}\neq\vec{k}'} -J_{\vec{k},\vec{k}'}\boldsymbol{\sigma}_{\vec{k}} \cdot \boldsymbol{\sigma}_{\vec{k}'},\quad J_{\vec{k},\vec{k}'} = \frac{1}{2N} \sum_{\alpha, \beta } \mathcal{V}^{\alpha\beta}_{\vec{k} - \vec{k}'} \phi_{\vec{k},\alpha}^* \phi_{\vec{k},\beta} \phi_{\vec{k}',\beta}^* \phi_{\vec{k}',\alpha}.
\end{align}

\subsection{Checkerboard model and mean-field solution}

We now specifically discuss the Hubbard interaction projected onto the degenerate ground state of the checkerboard lattice model introduced earlier in \equref{eq:HK_checkerboard_si}. Let us consider on-site and nearest-neighbors interactions which are given by 
\begin{subequations}
\begin{align}
    \Delta H &= H_{V} + H_{V'},\\
    H_{V} &= V \sum_{\vec{R},l} n_{\vec{R}+\vec{a}_l,\uparrow}
    n_{\vec{R}+\vec{a}_{l},\downarrow} = \frac{V}{2} \sum_{\vec{R},\sigma,\sigma'} : n_{\vec{R}+\vec{a}_l,\sigma}
    n_{\vec{R}+\vec{a}_{l},\sigma'} :\\
    H_{V'} &= \frac{V'}{4} \sum_{\vec{R},\sigma,\sigma^\prime} n_{\vec{R},\sigma}^{\pd} \sum_{j=0}^{3} n_{\vec{R}+C_{4z}^j\vec{a}_B,\sigma^\prime}.
\end{align}
\end{subequations}
Following the convention~\eqref{eq:fourier_transform_convention_si} of the Fourier transform, we can rewrite the interaction in $\vec{k}$-space in the form of~\eqref{eq:H_int} with the following interaction functions,
\begin{align}
    \mathcal{V}_{\vec{q}}^{AA} = \mathcal{V}_{\vec{q}}^{BB} = \frac{V}{2},\quad \mathcal{V}_{\vec{q}}^{AB} = \mathcal{V}_{\vec{q}}^{BA} = \frac{V'}{2} \cos\left(\frac{q_x}{2}\right) \cos\left(\frac{q_y}{2}\right).
\end{align}
Recalling that the lower-band wavefunctions of the non-interacting model are real~\eqref{eq:hk_eigenstates_si}, $\phi_{\vec{k},l} = U_{l,-}(\vec{k})$, we obtain the coupling constants of the effective spin-$1/2$ Heisenberg model (which is equivalent to the projected Hamiltonian $P\Delta H P$) as
\begin{align}
    J_{\vec{k},\vec{k}'} = \frac{V}{4N} \left[ \phi_{\vec{k},A}^2 \phi_{\vec{k}',A}^2 + \phi_{\vec{k},B}^2 \phi_{\vec{k}',B}^2 \right] + \frac{V'}{2N} \cos\left( \frac{k_x - k_x'}{2} \right) \cos\left( \frac{k_y - k_y'}{2} \right) \phi_{\vec{k},A} \phi_{\vec{k}',A} \phi_{\vec{k},B} \phi_{\vec{k}',B}.
\end{align}

We now solve this effective spin model in the mean-field approximation. If we further focus on collinear solutions, the problem of finding the MF ground state of the effective spin-model reduces to obtaining the ground state of the classical Ising model,
\begin{align}
    E[\{\sigma_{\vec{k}} \}] = 
   \sum_{\vec{k},\vec{k}' \in \mathcal{S}_1,\vec{k}\neq\vec{k}'} -J_{\vec{k},\vec{k}'}\sigma_{\vec{k}} \sigma_{\vec{k}'},\quad  \{ \sigma_{\vec{k}} \}^{\min} = \operatorname*{argmin}_{\{\sigma_{\vec{k}}=\pm\}} E[\{\sigma_{\vec{k}} \}].
\end{align}
From the spin-configuration $\{ \sigma_{\vec{k}} \}^{\min} $, we straightforwardly construct the corresponding fermionic state of the original model,
\begin{align}
    \ket{\Omega_{\mathrm{P+MF}}} = \ket{\{ \sigma_{\vec{k}} \}^{\min}} = \prod_{\vec{k}\in\mathcal{S}_1} f^{\dagger}_{\vec{k},\sigma_{\vec{k}}^{\min}} \ket{\mathcal{S}_2},
\end{align}
where we associate $\pm$ values of $\sigma_{\vec{k}}^{\min}$ with $\uparrow/\downarrow$ spin-projection of the fermions. This state is a collinear MF ground state of $P\Delta H P$.

\subsection{Construction of the one-shot MF Hamiltonian}

Having found the MF ground state $\ket{\Omega_{\mathrm{P + MF}}}$, we now want to demonstrate how the corresponding long-range order affects the spectral properties of the electrons. To do that, we herein construct the one-shot quadratic MF Hamiltonian based on the correlations of $\ket{\Omega_{\mathrm{P + MF}}}$. 

Let us first isolate the $\vec{k}$-local part of the Hamiltonian~\eqref{eq:H_int},
\begin{align}
    \begin{aligned}
        \Delta H = \frac{1}{N} \sum_{\vec{k},\vec{k}',\vec{q}} \mathcal{V}_{\vec{q}}^{\alpha\beta} : c^\dagger_{\vec{k}+\vec{q},\alpha} c^{\pd}_{\vec{k},\alpha} c^\dagger_{\vec{k}'-\vec{q},\beta} c^{\pd}_{\vec{k}',\beta} : &= \frac{1}{N} \mathcal{V}_{0}^{\alpha\beta} : c^\dagger_{\vec{k},\alpha} c^{\pd}_{\vec{k},\alpha} c^\dagger_{\vec{k},\beta} c^{\pd}_{\vec{k},\beta} : \\
        &+ \frac{1}{N} \sum_{\substack{\vec{k},\vec{k}',\vec{q} \\ (\vec{k},\vec{q})\neq (\vec{k}',\vec{0})}} \mathcal{V}_{\vec{q}}^{\alpha\beta} : c^\dagger_{\vec{k}+\vec{q},\alpha} c^{\pd}_{\vec{k},\alpha} c^\dagger_{\vec{k}'-\vec{q},\beta} c^{\pd}_{\vec{k}',\beta}:,
    \end{aligned}
\end{align}
where we suppressed the spin index for notations simplicity.
Imposing translational invariance, we perform the MF decomposition (Hartree-Fock) of the second term. Then the MF version of the Hamiltonian $\Delta H$ reads
\begin{subequations}
    \begin{align}
    \Delta H_{\mathrm{MF}} &= \frac{1}{N} \mathcal{V}_{0}^{\alpha\beta} : c^\dagger_{\vec{k},\alpha} c^{\pd}_{\vec{k},\alpha} c^\dagger_{\vec{k},\beta} c^{\pd}_{\vec{k},\beta}: \nonumber \\
    &+ \frac{1}{N} \sum_{\vec{k},\vec{k}',\vec{k}\neq\vec{k}'} \mathcal{V}_0^{\alpha\beta} \left[ \hat{n}_{\vec{k}',\alpha} \braket{\hat{n}_{\vec{k},\beta}} +  \braket{\hat{n}_{\vec{k}',\alpha}} \hat{n}_{\vec{k},\beta} - \braket{\hat{n}_{\vec{k}',\alpha}} \braket{\hat{n}_{\vec{k},\beta}} \right]\\
    &- \frac{1}{2N} \sum_{\vec{k},\vec{k}',\vec{k}\neq\vec{k}'} \mathcal{V}_{\vec{k}'-\vec{k}}^{\alpha\beta} \left[ \braket{C^{\alpha\beta}_{\vec{k}'}} C^{\beta\alpha}_{\vec{k}} + C^{\alpha\beta}_{\vec{k}'} \braket{C^{\beta\alpha}_{\vec{k}}} - \braket{C^{\alpha\beta}_{\vec{k}'}} \braket{C^{\beta\alpha}_{\vec{k}}}  \right]\\
    &-\frac{2}{N} \sum_{\vec{k},\vec{k}',\vec{k}\neq\vec{k}'} \mathcal{V}_{\vec{k}'-\vec{k}}^{\alpha\beta} \left[ \braket{\vec{S}^{\alpha\beta}_{\vec{k}'}} \cdot \vec{S}^{\beta\alpha}_{\vec{k}} + \vec{S}^{\alpha\beta}_{\vec{k}'} \cdot \braket{\vec{S}^{\beta\alpha}_{\vec{k}}} - \braket{\vec{S}^{\alpha\beta}_{\vec{k}'}} \cdot \braket{\vec{S}^{\beta\alpha}_{\vec{k}}}  \right].
\end{align}    
\end{subequations}
This Hamiltonian is consistent with our prior projection and MF calculations. Let us elaborate on what we mean by consistency. The projection and MF decomposition do not generally commute, nonetheless, the expectation value of the constructed Hamiltonian in the state $\ket{\Omega_{\mathrm{P+MF}}}$ coincides with the energy of this state obtained $\textit{after}$ projection and MF,
\begin{align}
    \braket{\Omega_{\mathrm{P+MF}} | \Delta H_{\mathrm{MF}}[\ket{\Omega_{\mathrm{P+MF}}}] | \Omega_{\mathrm{P+MF}}} = E_{\mathrm{P+MF}}.
\end{align}
An exact treatment of the $\vec{k}$-local part is required because the obtained MF state is the product state only at the $\vec{k}$-level, $\ket{\Omega_{\mathrm{P+MF}}} = \prod_{\vec{k}} \ket{\Psi_{\vec{k}}}$, while each $\ket{\Psi_{\vec{k}}}$ can generally exhibit orbital mixing/entanglement in the doubly occupied region $\mathcal{S}_2$.

Our primary reason for constructing the MF Hamiltonian is to demonstrate the spin-splitting in the spectral function. Hence, we neglect the spin-diagonal band renormalization terms, and consider only the SU(2) breaking part of the MF Hamiltonian. In the main text, we use the following spin-splitting Hamiltonian constructed from the MF solution $\ket{\Omega_{\mathrm{P+MF}}}$,
\begin{align}
    \begin{aligned}
        \Delta H_{\mathrm{MF}}^{\text{spin-splitting}}[ \ket{\Omega_{\mathrm{P+MF}}} ] = &- \frac{2}{N} \sum_{\vec{k}} \mathcal{V}_{\vec{k}-\vec{k}'}^{\alpha\beta} \vec{S}_{\vec{k}}^{\alpha\beta} \cdot \sum_{\vec{k}'\in\mathcal{S}_1,\vec{k}'\neq\vec{k}} \Braket{ \Omega_{\mathrm{P+MF}} |  \vec{S}_{\vec{k}'}^{\beta\alpha} | \Omega_{\mathrm{P+MF}}} \\
        &- \frac{2}{N} \sum_{\vec{k}} \mathcal{V}_{\vec{k}'-\vec{k}}^{\alpha\beta} \vec{S}_{\vec{k}}^{\beta\alpha} \cdot \sum_{\vec{k}'\in\mathcal{S}_1,\vec{k}'\neq\vec{k}} \Braket{\Omega_{\mathrm{P+MF}} | \vec{S}_{\vec{k}'}^{\alpha\beta} | \Omega_{\mathrm{P+MF}}}.
    \end{aligned}
\label{eq:SpinSplitting}
\end{align}

\section{Expression for the static spin correlator}

In this section, we consider the HK model with an additional interaction term $\delta H$, that still allows the exact diagonalization of the full Hamiltonian, $H = H_{\mathrm{HK}} + \delta H$. For simplicity, let us first focus on the one-site HK model. We are interested in the interactions that fully or partially remove the degeneracy of the HK ground state. One example of such an interaction is
\begin{align}
    \delta H_{2S} &= -J \sum_{\vec{k}} \vec{S}_{\vec{k}} \cdot \vec{S}_{-\vec{k}}, \label{eq:k_minus_k_perturbation_si}
\end{align}
which fully lifts the degeneracy when $J < 0$. The ground state in the singly occupied region then consists of decoupled singlets formed by electrons at $\vec{k}$ and $-\vec{k}$,
\begin{align}
    \ket{\mathrm{GS}} = \prod_{\vec{k}\in\mathcal{S}_1, k_x>0} \frac{1}{\sqrt{2}} \left[ c^\dagger_{\vec{k},\uparrow} c^{\dagger}_{-\vec{k},\downarrow} - c^\dagger_{\vec{k},\downarrow} c^{\dagger}_{-\vec{k},\uparrow} \right] \prod_{\vec{k}'\in\mathcal{S}_2} c^\dagger_{\vec{k}',\uparrow} c^\dagger_{\vec{k}',\downarrow} \ket{0}.
\end{align}
Note that $\delta H$ modifies the singly and doubly occupied regions. The subsequent calculations are based on the assumption that $\hat{n}_{\vec{k}} = \sum_{\sigma}\hat{n}_{\vec{k},\sigma}$ is conserved. It means that we either treat $\delta H$ as a perturbation (by diagonalizing $P \delta H P$) or we diagonalize the full Hamiltonian $H_{\mathrm{HK}} + \delta H$, and require that $[\delta H, \hat{n}_{\vec{k}}] = 0$. Finally, we denote the set of momenta that are coupled by the interaction $\delta H$ to a given $\vec{k}$ by $C(\vec{k})$, which includes $\vec{k}$ as well. For example, the interaction~\eqref{eq:k_minus_k_perturbation_si} couples only two momenta, $\vec{k}$ and $-\vec{k}$. Momenta from different clusters $C$ are uncorrelated.

\subsection{Spin correlator}

We want to understand how the additional interaction $\delta H$ affect the spin correlations,
\begin{align}
    \Braket{S^z(\vec{0}) S^z(\mathbf{R})} = \frac{1}{4N^2} \sum_{\vec{k}_1,\vec{k}_2,\vec{k},\vec{q}} \Braket{c^\dagger_{\vec{k}_1}\sigma_z c^{\pd}_{\vec{k}_2} c^\dagger_{\vec{k}+\vec{q}}\sigma_z c^{\pd}_{\vec{k}}} e^{-i\vec{q}\vec{R}}.
\end{align}
Since $\hat{n}_{\vec{k}}$ is conserved, only the terms corresponding to the following combination of momenta give non-vanishing contribution,
\begin{subequations}
\begin{alignat}{3}
    \vec{q} &= \vec{0},\quad &&\vec{\vec{k}_1} &&= \vec{k}_2;\\
    \vec{k}+\vec{q} &= \vec{k}_2,\quad && \vec{k}_1 &&= \vec{k}.
\end{alignat}
\end{subequations}
Therefore,
\begin{align}
    \Braket{S^z(\vec{0}) S^z(\mathbf{R})} = \frac{1}{N^2} \sum_{\vec{k},\vec{k}'} \Braket{S^z_{\vec{k}} S^z_{\vec{k}'}} + \frac{1}{4N^2} \sum_{\vec{k},\vec{k}',\vec{k}\neq\vec{k}'} \braket{c^\dagger_{\vec{k}} \sigma_z c^{\pd}_{\vec{k}'} c^\dagger_{\vec{k}'}\sigma_z c^{\pd}_{\vec{k}}} e^{-i(\vec{k}' - \vec{k})\vec{R}}. \label{eq:correlator_main}
\end{align}
It is useful to split the summation over $\mathbf{k}'$ into correlated and uncorrelated parts. The first term then reads
\begin{align}
    \sum_{\vec{k},\vec{k}'} \braket{S^z_{\vec{k}} S^z_{\vec{k}'}} &= \sum_{\vec{k};\vec{k}'\in C(\vec{k})} \Braket{S^z_{\vec{k}} S^z_{\vec{k}'}}  + \sum_{\vec{k}; \vec{k}'\in \mathrm{BZ} \setminus  C(\vec{k})} \Braket{S^z_{\vec{k}} S^z_{\vec{k}'}} \nonumber \\
    &= \sum_{\vec{k};\vec{k}'\in C(\vec{k})} \Braket{S^z_{\vec{k}} S^z_{\vec{k}'}} + \sum_{\vec{k}; \vec{k}'\in \mathrm{BZ} \setminus  C(\vec{k})} \Braket{S^z_{\vec{k}}} \Braket{S^z_{\vec{k}'}} \nonumber  \\
    &= \sum_{\vec{k};\vec{k}'\in C(\vec{k})} \Braket{S^z_{\vec{k}} S^z_{\vec{k}'}} \nonumber \\
    &= \frac{1}{3} \sum_{\vec{k};\vec{k}'\in C(\vec{k})} \Braket{\vec{S}_{\vec{k}} \cdot \vec{S}_{\vec{k}'}}, \label{eq:spin_spin_part}
\end{align}
where we exploited SU(2) symmetry, $\braket{S_{\vec{k}}^j} = 0$, $\braket{S_{\vec{k}}^j S_{\vec{k}'}^{j}} = 1/3 \Braket{\vec{S}_{\vec{k}} \cdot \vec{S}_{\vec{k}'}}$. 

We now rewrite the expectation value in the second term of Eq.~\eqref{eq:correlator_main},
\begin{align}
    \braket{c^\dagger_{\vec{k}} \sigma_z c^{\pd}_{\vec{k}'} c^\dagger_{\vec{k}'}\sigma_z c^{\pd}_{\vec{k}}} &= - \sum_{\sigma} \Braket{\hat{n}_{\vec{k},\sigma} \hat{n}_{\vec{k}',\sigma}} + \braket{n_{\vec{k}}} + \sum_{\sigma} \braket{c^{\dagger}_{\vec{k},\sigma} c^{\pd}_{\vec{k},\overline{\sigma}} c^{\dagger}_{\vec{k}',\overline{\sigma}} c^{\pd}_{\vec{k}',\sigma}}  \nonumber \\
    &= -\frac{1}{2} \braket{\hat{n}_{\vec{k}} \hat{n}_{\vec{k}'}} + \braket{n_{\vec{k}}} - 2 \Braket{S^z_{\vec{k}} S^z_{\vec{k}'}} + 2 \Braket{S^x_{\vec{k}} S^x_{\vec{k}'}} + 2 \Braket{S^y_{\vec{k}} S^y_{\vec{k}'}}  \nonumber \\
    &= -\frac{1}{2}  \braket{\hat{n}_{\vec{k}}} \braket{\hat{n}_{\vec{k}'}} + \braket{n_{\vec{k}}} + \frac{2}{3} \Braket{\vec{S}_{\vec{k}} \cdot \vec{S}_{\vec{k}'}}.
\end{align}
Using the same splitting as in~\eqref{eq:spin_spin_part} and again exploiting SU(2), we finally obtain
\begin{align}
    \Braket{\vec{S}(\vec{0}) \cdot \vec{S}(\vec{R})} = & - \frac{3}{8N^2} \left| \sum_{\vec{k}} \braket{\hat{n}_{\vec{k}}} e^{i\mathbf{kR}} \right|^2 + \delta_{\vec{R},\vec{0}} \frac{3}{4N} \sum_{\vec{k}} \braket{\hat{n}_{\vec{k}}} \nonumber \\
    & + \frac{1}{2N^2} \sum_{\vec{k},\vec{k}'\in C(\vec{k})} \Braket{\vec{S}_{\vec{k}} \cdot \vec{S}_{\vec{k}'}} \biggl\{ 2 + \cos[(\vec{k}' - \vec{k})\vec{R}] \biggr\} - \frac{3}{4N^2}\sum_{\vec{k}\in\mathcal{S}_1} 1.\label{eq:s0sR_corr}
\end{align}
Thus, we can clearly see that the all the information about the perturbation $\delta H$ is encoded in the spin-spin correlations $\Braket{\vec{S}_{\vec{k}} \cdot \vec{S}_{\vec{k}'}}$. If $\delta H$ couples finite number (independent of the system size $N$) of momenta (e.g. the interaction~\eqref{eq:k_minus_k_perturbation_si} couples two momenta), the interaction-induced correlation vanish in the thermodynamic limit, and spin-correlator is fully determined by the fermionic statistics,
\begin{align}
    \Braket{\vec{S}(\vec{0}) \cdot \vec{S}(\vec{R})} = -\frac{3}{8} \left| \frac{1}{(2\pi)^d} \int_{\mathrm{BZ}} \diff^d \vec{k} \braket{\hat{n}_{\vec{k}}} e^{i\mathbf{kR}} \right|^2 + \delta(\vec{R}) \frac{3}{4} \frac{1}{(2\pi)^d} \int_{\mathrm{BZ}} \diff^d \vec{k} \braket{\hat{n}_{\vec{k}}}.
\end{align}
This expression coincides with the result for the Fermi sea, which is the consequence of the fact that the HK model inherits a lot of properties from non-interacting systems.

\subsection{Static spin structure factor}

From \equref{eq:s0sR_corr}, we can straightforwardly obtain the static spin-structure factor,
\begin{align}
    S({\vec{q}}) & = \sum_{\vec{R}} \Braket{\vec{S}(\vec{0}) \cdot \vec{S}(\vec{R})} e^{i\vec{qR}}\nonumber\\
    &= -\frac{3}{8N} \sum_{\vec{k}} n_{\vec{k}} n_{\vec{k}+\vec{q}} + \frac{3}{4N} \sum_{\vec{k}} n_{\vec{k}} + \frac{1}{N} \sum_{\vec{k},\vec{k}'\in C(\vec{k})} \Braket{\vec{S}_{\vec{k}} \cdot \vec{S}_{\vec{k}'}} \left[ \delta_{\vec{q},\vec{0}} + \frac{1}{2} \delta_{\vec{k}',\vec{k}+\vec{q}} \right] - \frac{3}{4N} \delta_{\vec{q},\vec{0}} \sum_{\vec{k}\in\mathcal{S}_1} 1\nonumber\\
    &= \frac{3}{8 N} \sum_{\vec{k}} n_{\vec{k}}(2 - n_{\vec{k}+\vec{q}}) + \frac{1}{N} \sum_{\vec{k},\vec{k}'\in C(\vec{k})} \Braket{\vec{S}_{\vec{k}} \cdot \vec{S}_{\vec{k}'}} \left[ \delta_{\vec{q},\vec{0}} + \frac{1}{2} \delta_{\vec{k}',\vec{k}+\vec{q}} \right] - \frac{3}{4N} \delta_{\vec{q},\vec{0}} \sum_{\vec{k}\in\mathcal{S}_1} 1.
\end{align}

The ferromagnetic structure factor then reads
\begin{align}
    S(\vec{q}=\vec{0}) &= \frac{1}{N} \sum_{\vec{k};\vec{k}'\in C(\vec{k})} \Braket{\vec{S}_{\vec{k}} \cdot \vec{S}_{\vec{k}'}}= \frac{1}{N}\sum_{C} \sum_{\vec{k},\vec{k}'\in C} \braket{\vec{S}_{\vec{k}} \cdot \vec{S}_{\vec{k}'}}\nonumber \\
    &= \frac{1}{N} \sum_{C}\Braket{\biggl(\sum_{\vec{k}\in C} \vec{S}_{\vec{k}} \biggr)^2},
\end{align}
where $\sum_C$ denotes the summation over clusters of coupled momenta. We can see that depending on the perturbation we can modify the ferromagnetic correlations, even enhance them or fully suppress. For example, the four-spin perturbation,
\begin{align}
    \delta H = - J\sum_{\vec{k}} \sum_{j=0}^3 \vec{S}_{C_{4z}^j\vec{k}} \cdot \vec{S}_{C_{4z}^{j+1}\vec{k}},
\end{align}
removes the degeneracy and results in the unique ground state when $J < 0$. The ground state is now a singlet state which fully suppresses the ferromagnetic structure factor. Note that $J > 0$ would enhance $S(\vec{q}=\vec{0})$.

The finite momentum spin structure factor is given by
\begin{align}
    S(\vec{q}\neq \vec{0}) = \frac{1}{2N}\sum_{\vec{k}\in\mathcal{S}_1} \mathbf{1}_{\mathcal{S}_1}(\vec{k}+\vec{q}) \left[ \frac{3}{4} + \mathbf{1}_{C(\vec{k})}(\vec{k}+\vec{q}) \braket{\vec{S}_{\vec{k}} \cdot \vec{S}_{\vec{k}+\vec{q}}} \right], \label{FiniteqForm}
\end{align}
where $\mathbf{1}_{A}(x)$ is the indicator function. Analogously to the ferromagnetic correlations, the perturbation $\delta H = -J \sum_{\vec{k}} \vec{S}_{\vec{k}} \cdot \vec{S}_{\vec{k}+\vec{Q}} $ with $\vec{Q} = (\pi,\pi)$ fully suppresses the antiferromagnetic correlations $S(\vec{Q})$ when $J<0$. This interaction, however, does not lift the degeneracy for an arbitrary filling.

\subsection{Multiorbital case}

We now generalize the calculations of the spin-spin correlations to the multi-orbital HK model of the form~\eqref{eq:multiorbital_HK} with an additional perturbation that modifies spin-spin correlations of the unperturbed HK ground state while conserving SU(2) symmetry and the number of particles at each momentum $\vec{k}$. It is instructive to consider the following spin-spin perturbation
\begin{align}
    \delta H = - \sum_{C} \sum_{\vec{k},\vec{k}'\in C}  \sum_{\alpha,\beta,\alpha',\beta'} \tilde{J}_{\vec{k},\vec{k}'}^{\alpha\beta\alpha'\beta'}\, \vec{S}_{\vec{k}}^{\alpha\beta} \cdot \vec{S}_{\vec{k}'}^{\alpha'\beta'},
\end{align}
where the summation $\sum_{C}$ goes over non-overlapping finite-size clusters $C$ of coupled momenta.

The spin-spin correlator reads
\begin{align}
    \Braket{S^{\alpha\beta}_z(\vec{0}) S_{z}^{\alpha'\beta'}(\vec{R})} 
    &= \frac{1}{N} \sum_{\vec{k},\vec{k}'} \Braket{S_{\vec{k},z}^{\alpha\beta} S_{\vec{k}',z}^{\alpha'\beta'}} + \frac{1}{4N^2} \sum_{\vec{k},\vec{k}',\vec{k}\neq\vec{k}'}  \Braket{c^\dagger_{\vec{k},\alpha} \sigma_z c^{\pd}_{\vec{k}',\beta} c^\dagger_{\vec{k}',\alpha'} \sigma_z c^{\pd}_{\vec{k},\beta'}} e^{-i(\vec{k}' - \vec{k})\vec{R}},
\end{align}
where we use the following convention for the Fourier transform, $c_{\vec{R},\alpha} = N^{-1/2} \sum_{\vec{k}} e^{i\vec{k} \vec{R}} c_{\vec{k},\alpha}$. The first term can be again split into correlated and uncorrelated parts. The uncorrelated part vanishes due to SU(2) symmetry, hence,
\begin{align}
    \Braket{S_{\vec{k},z}^{\alpha\beta} S_{\vec{k}',z}^{\alpha'\beta'}} &= \mathbf{1}_{C(\vec{k})}(\vec{k}') \Braket{S_{\vec{k},z}^{\alpha\beta} S_{\vec{k}',z}^{\alpha'\beta'}} + \left(1 - \mathbf{1}_{ C(\vec{k})}(\vec{k}') \right) \Braket{S_{\vec{k},z}^{\alpha\beta}} \Braket{S_{\vec{k}',z}^{\alpha'\beta'}}\nonumber\\
    &= \mathbf{1}_{C(\vec{k})}(\vec{k}') \Braket{S_{\vec{k},z}^{\alpha\beta} S_{\vec{k}',z}^{\alpha'\beta'}}.
\end{align}
The second term can be rewritten in the convenient way,
\begin{align}
    c^\dagger_{\vec{k},\alpha} \sigma_z c^{\pd}_{\vec{k}',\beta} c^\dagger_{\vec{k}',\alpha'} \sigma_z c^{\pd}_{\vec{k},\beta'} &= \sum_{\sigma} \left[ \delta_{\alpha',\beta} c^\dagger_{\vec{k},\alpha,\sigma} c^{\pd}_{\vec{k},\beta',\sigma} - c^\dagger_{\vec{k},\alpha,\sigma}c^{\pd}_{\vec{k},\beta',\sigma} c^\dagger_{\vec{k}',\alpha',\sigma} c^{\pd}_{\vec{k}',\beta,\sigma} + c^\dagger_{\vec{k},\alpha,\sigma}c^{\pd}_{\vec{k},\beta',\overline{\sigma}} c^\dagger_{\vec{k}',\alpha',\overline{\sigma}} c^{\pd}_{\vec{k}',\beta,\sigma} \right] \nonumber \\
    &= \delta_{\alpha',\beta} C^{\alpha\beta'}_{\vec{k}} - \frac{1}{2} C^{\alpha\beta'}_{\vec{k}} C^{\alpha'\beta}_{\vec{k}'} - 2 S^{\alpha\beta'}_{\vec{k},z} S^{\alpha'\beta}_{\vec{k}',z} + 2 S^{\alpha\beta'}_{\vec{k},x} S^{\alpha'\beta}_{\vec{k}',x} + 2 S^{\alpha\beta'}_{\vec{k},y} S^{\alpha'\beta}_{\vec{k}',y}.
\end{align}
Due to the SU(2) symmetry, we get
\begin{align}
    \Braket{c^\dagger_{\vec{k},\alpha} \sigma_z c^{\pd}_{\vec{k}',\beta} c^\dagger_{\vec{k}',\alpha'} \sigma_z c^{\pd}_{\vec{k},\beta'}} & = \delta_{\alpha',\beta} \Braket{C^{\alpha\beta'}_{\vec{k}}} - \frac{1}{2} \Braket{C^{\alpha\beta'}_{\vec{k}} C^{\alpha'\beta}_{\vec{k}'}} + \frac{2}{3} \Braket{\vec{S}_{\vec{k}}^{\alpha\beta'} \cdot \vec{S}_{\vec{k}'}^{\alpha'\beta}}\nonumber \\
    &=  \begin{aligned}[t]
        \delta_{\alpha',\beta} \Braket{C^{\alpha\beta'}_{\vec{k}}} & + \mathbf{1}_{C(\vec{k})}(\vec{k}')\left[ \frac{2}{3}\Braket{\vec{S}_{\vec{k}}^{\alpha\beta'} \cdot \vec{S}_{\vec{k}'}^{\alpha'\beta}} - \frac{1}{2} \Braket{C^{\alpha\beta'}_{\vec{k}} C^{\alpha'\beta}_{\vec{k}'}}  \right] \\
        &- \left(1 - \mathbf{1}_{ C(\vec{k})}(\vec{k}') \right) \frac{1}{2} \Braket{C^{\alpha\beta'}_{\vec{k}}} \Braket{C^{\alpha'\beta}_{\vec{k}'}}.
    \end{aligned}
\end{align}

Thus,
\begin{align}
    \begin{aligned}
        \Braket{\vec{S}^{\alpha\beta}(\vec{0}) \cdot \vec{S}^{\alpha'\beta'}(\vec{R})} &=  \delta_{\alpha',\beta} \frac{3}{4N} \sum_{\vec{k}}  \Braket{C^{\alpha\beta'}_{\vec{k}}} \left[ \delta_{\vec{R},\vec{0}} - \frac{1}{N} \right] -  \frac{3}{8N^2} \sum_{\vec{k};\vec{k}'\not\in C(\vec{k})}  \Braket{C^{\alpha\beta'}_{\vec{k}}} \Braket{C^{\alpha'\beta}_{\vec{k}'}} e^{i(\vec{k} - \vec{k}')\vec{R}}\\
        &+\frac{1}{N^2} \sum_{\vec{k};\vec{k}'\in C(\vec{k})} \Braket{\vec{S}_{\vec{k}}^{\alpha\beta} \cdot \vec{S}_{\vec{k}'}^{\alpha'\beta'}} \\
        &+ \frac{1}{2N^2} \sum_{\substack{\vec{k};\vec{k}'\in C(\vec{k}) \\ \vec{k}\neq\vec{k}' }} \left[ \Braket{\vec{S}_{\vec{k}}^{\alpha\beta'} \cdot \vec{S}_{\vec{k}'}^{\alpha'\beta}} - \frac{3}{4} \Braket{C^{\alpha\beta'}_{\vec{k}} C^{\alpha'\beta}_{\vec{k}'}}  \right] e^{i(\vec{k} - \vec{k}')\vec{R}}.
    \end{aligned}
\end{align}

Taking the Fourier transform, we find the static spin structure factor, $S^{\alpha\beta\alpha'\beta'}(\vec{q}) = \sum_{\vec{R}} \Braket{\vec{S}^{\alpha\beta}(\vec{0}) \cdot \vec{S}^{\alpha'\beta'}(\vec{R})} e^{-i\vec{q}\vec{R}}$. As in the single-orbital case, for $\vec{q}=\vec{0}$, it is again determined solely by the perturbation-induced spin-correlations,
\begin{align}
    S^{\alpha\beta\alpha'\beta'}(\vec{q}=\vec{0}) = \frac{1}{N} \sum_{C} \Braket{\sum_{\vec{k}\in C} \vec{S}_{\vec{k}}^{\alpha\beta} \cdot \sum_{\vec{k}'\in C} \vec{S}_{\vec{k}'}^{\alpha'\beta'}}.
\end{align}
The finite-momentum spin-structure factor reads
\begin{align}
    \begin{aligned}
        S^{\alpha\beta\alpha'\beta'} (\vec{q} \neq \vec{0}) &= \delta_{\alpha',\beta} \frac{3}{4N} \sum_{\vec{k}}  \Braket{C^{\alpha\beta'}_{\vec{k}}} - \frac{3}{8N} \sum_{\vec{k}} \left[1 - \mathbf{1}_{C(\vec{k})}(\vec{k}+\vec{q}) \right]  \Braket{C^{\alpha\beta'}_{\vec{k}}} \Braket{C^{\alpha'\beta}_{\vec{k}+\vec{q}}}\\
        &+ \frac{1}{2N} \sum_{\vec{k}} \mathbf{1}_{C(\vec{k})}(\vec{k}+\vec{q}) \left[ \Braket{\vec{S}_{\vec{k}}^{\alpha\beta'} \cdot \vec{S}_{\vec{k}+\vec{q}}^{\alpha'\beta}} - \frac{3}{4} \Braket{C^{\alpha\beta'}_{\vec{k}} C^{\alpha'\beta}_{\vec{k}+\vec{q}}}  \right].
    \end{aligned}
\end{align}
Although the multiorbital nature of this expression makes it generally difficult to analyze, we clearly see that the perturbation $\delta H$ induces spin-correlations at finite momentum $\vec{q}$ only if there exists $\vec{k}$ such that both $\vec{k}$ and $\vec{k}+\vec{q}$ lie in the same cluster $C$.

\section{Further spectral functions}

\begin{figure}[h]
    \centering
    \includegraphics[width=\linewidth]{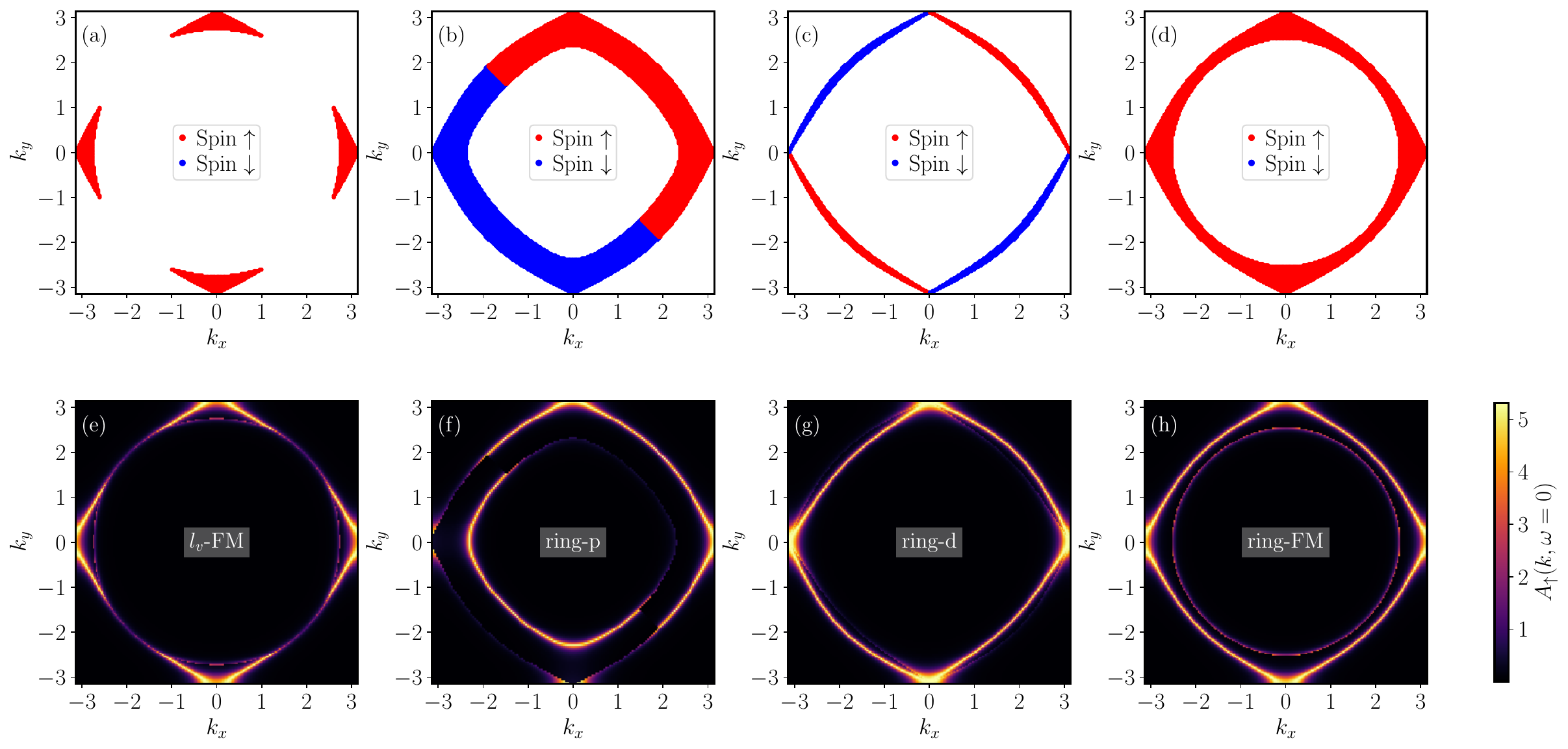}
    \caption{Different spin configurations (a-d) of ring and $l_v$ $\mathcal{S}_1$, that are obtained from the minimization, and their corresponding spin up spectral functions $A_\uparrow (k,\omega=0)$ (e-h). Combined with the spectral functions shown in the \textit{long range order} section of the main text, the shown spectral functions give examples for all phases in the phase diagram Fig. 3c.}
    \label{fig:SpecAppend}
\end{figure}

\begin{figure}
    \centering
    \includegraphics[width=0.75\linewidth]{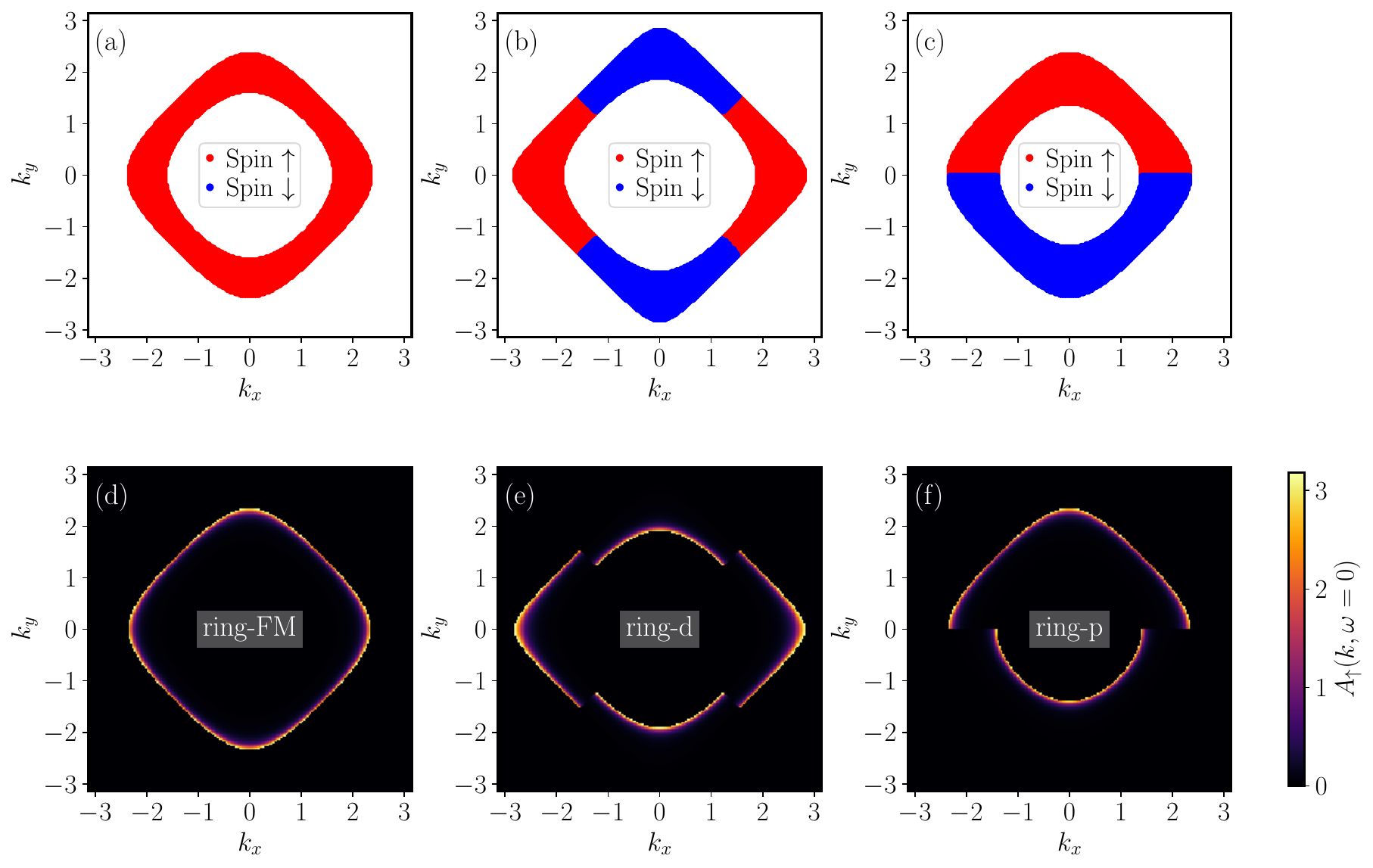}
    \caption{Spin configurations of the three magnetic phases in the ring regime of the single-orbtial HK model and their corresponding spin up spectral functions $A_\uparrow (k,\omega=0)$. The single-orbital HK model only allows ring and disk $\mathcal{S}_1$. Therefore, the spectral functions shown constitute a comprehensive set of non-trivial spectral functions.}
    \label{fig:1Orbit_Append}
\end{figure}

Following the calculations described in the \textit{long-range order} section of the main text, we compute the spin-split spectral functions for the cases: $l_v$-FM, ring-$p$, ring-$d$ and ring-FM. They are shown in \figref{fig:SpecAppend}(e-h). We also show the corresponding spin configurations that we obtain via minimization in $\mathcal{S}_1$. They are shown in \figref{fig:SpecAppend}(a-d). The interactions $U$ and $U^\prime$ for the calculations are chosen as:
\begin{center}
\begin{tabular}{l l l }
$\bullet$ $l_v$-FM: & $U=1.0$, &$U^\prime = -1.0$ \\
$\bullet$ ring-$p$: & $U=0.2$, &$U^\prime = 2.6$ \\
$\bullet$ ring-$d$: & $U=0.003$, &$U^\prime = 0.5$ \\
$\bullet$ ring-FM:  & $U=1.0$, &$U^\prime = -0.1$~.
\end{tabular}
\end{center}
All shown spectral functions (in main text and appendix) share the same chemical potential $\mu =-1.0$ and next nearest-neighbor hopping $t_2=0.5$.
We use a system with size $200 \times 200$ and zero temperature $T=0$.

The ring spin up spectral functions $A_\uparrow (k,\omega=0)$ in \figref{fig:SpecAppend} show the symmetries discussed in the main text. The respective spin down spectral functions $A_\downarrow (k,\omega=0)$ of the $p$-wave and $d$-wave altermagnet can be obtained via rotation of $A_\uparrow (k,\omega=0)$. For $p$-wave this is a rotation by $\pi$ and for $d$-wave this is a rotation by $\frac{\pi}{2}$.
\\
\\
For the single-orbital HK model
\begin{equation}
    H_{\text{HK},1} = \sum_{\boldsymbol{k},\sigma} (\xi_{\boldsymbol{k}} - \mu) \hat{n}_{\boldsymbol{k},\sigma} + U \sum_{\boldsymbol{k}} \hat{n}_{\boldsymbol{k},\uparrow} \hat{n}_{\boldsymbol{k},\downarrow} + U^\prime \sum_{\boldsymbol{k},\sigma,\sigma^\prime} \hat{n}_{\boldsymbol{k},\sigma} \hat{n}_{\boldsymbol{k},\sigma^\prime} 
\end{equation}
we follow the calculations of the two-orbital HK model, that are described in appendix B. We find the same magnetic phases in the single-orbital model that where already present in the two-orbital model. The spin-up spectral functions $A_\uparrow(\boldsymbol{k},\omega=0)$ of the ring-FM, ring-$d$ and ring-$p$ configuration are shown in \figref{fig:1Orbit_Append}d to f. For this we use the equivalent of \equref{eq:SpinSplitting}. We also show the corresponding spin configurations in $\mathcal{S}_1$ that are obtained via minimization \figref{fig:1Orbit_Append}(a-c). The corresponding spin-down spectral functions behave as expected and can be generated with rotations of $\frac{\pi}{2}$/$\pi$ in the $p/d$-wave case. For the single-orbital spectral functions we chose a system size of $200\times 200$ and temperature $T=0$, chemical potential $\mu$ and interactions $U$ and $U^\prime$  are varied to obtain the different configurations.

\end{appendix}

\end{document}